\documentclass{article}
\usepackage{natbib}
\usepackage{epubtk}    
\usepackage{booktabs}  
\usepackage{graphicx}  

\showlistoftablesfalse

\begin{document}

\title{Solar and Stellar Photospheric Abundances}

\author{\epubtkAuthorData{Carlos Allende Prieto}{%
Instituto de Astrof\'{\i}sica de Canarias \\
C/ V\'{\i}a L\'actea S/N, E-38200 La Laguna, Tenerife, Spain}{%
callende@iac.es}{%
http://hebe.as.utexas.edu}%
}

\date{}
\maketitle

\begin{abstract}
The determination of photospheric abundances in late-type stars 
from spectroscopic observations is a well-established field, built on
solid theoretical foundations. Improving those foundations to refine
the accuracy of the inferred abundances has proven challenging, 
but progress has been made. In parallel, 
developments on instrumentation, chiefly regarding multi-object spectroscopy, 
have been spectacular, and a number of projects are collecting 
large numbers of observations for stars across the Milky Way and
nearby galaxies, promising important advances in our understanding of galaxy 
formation and evolution. After providing a brief description of the basic physics
and input data involved in the analysis of stellar spectra, a review is made 
of the analysis steps, and the available tools to cope with large 
observational efforts. The paper closes with a quick overview of 
relevant ongoing and planned spectroscopic surveys, and highlights
of recent research on photospheric abundances.
\end{abstract}

\epubtkKeywords{Abundances, Stellar atmospheres, Stellar spectra}

\newpage
\tableofcontents

\newpage

\section{Introduction}
\label{intro}


Stars drive the chemical evolution of the universe. Starting from hydrogen, 
stars build up heavier elements, first by nuclear fusion in their cores, 
and later by neutron bombardment in the final stages of their lives.
A fraction of the gas processed in stars makes it back to the interstellar
medium, where it can again collapse under gravity and form new stars,
repeating the cycle.

This picture is complicated by the fact that the stellar-processed gas mixes
with unpolluted (or less polluted) gas, diluting the \emph{metals}%
\epubtkFootnote{We talk about metals when referring 
to any element heavier than helium.}
available at any 
given location. Far from being a closed box, the Milky Way strips stars 
and gas from its immediate neighbors. Depending mainly on their mass, 
stars of different masses produce 
(and sometimes destroy) nuclei of different elements, but their 
nucleosynthetic yields also depend 
on other parameters such as chemical composition. 
Most heavy elements are actually produced 
during the final stages of the lives of stars, in nuclear reactions 
driven by supernovae explosions. 
Mixing within stars blurs and modifies the radial stratification 
predicted by simple, spherically symmetric, models of stars. 
In addition, close binaries lead to far 
more complex stellar evolution scenarios where the 
events expected for an isolated star may happen sooner, or never take place.

The formation and chemical evolution of the Galaxy is therefore 
closely entangled with the problem of stellar structure and evolution, 
which includes energy production and nucleosynthesis in stars. 
Observationally, we see the progressive enrichment in 
metals of the Milky Way in the chemical compositions of intermediate 
and low-mass stars that were born at different times. The chemical mixtures 
found in the surfaces of 
main-sequence stars, essentially undisturbed by the nuclear 
reactions going on deep in the interior, sample the chemistry of the gas 
in the interstellar medium 
from which they formed. Stars preserve those samples for us to study. 

Reading the chemical patterns frozen in the stellar surfaces 
involves solving a different problem, that of the structure of 
the outermost layers of a star.
Part of the energy initially produced in the nuclear reactions 
 in the interior of a star escapes immediately in the 
form of neutrinos; this amounts to $\sim$~2\% in the solar case. 
The rest diffuses outwards slowly, finally reaching the surface 
of the star where photons decouple from matter. The 
distribution of escaping photons reflects the state of the 
gas in the stellar atmosphere. Conversely, the radiation field 
has a profound influence on the physical structure of the 
atmosphere through its effect on the energy balance.

To model a stellar atmosphere we need to know how much 
energy flows through it, $\sigma T_{\mathrm{eff}}^4$, where $\sigma$ is
the Stefan--Boltzmann constant, the surface gravity of the star, 
$g = G M/R^2$, and its chemical 
composition. The observed spectrum constrains these parameters, 
but sometimes there is additional information available. 
For example, if we know the parallax of a star 
and its angular diameter, we can readily compute its radius. 

\epubtkImage{}{%
  \begin{figure}[htb]
    \centerline{\includegraphics[width=0.9\textwidth]{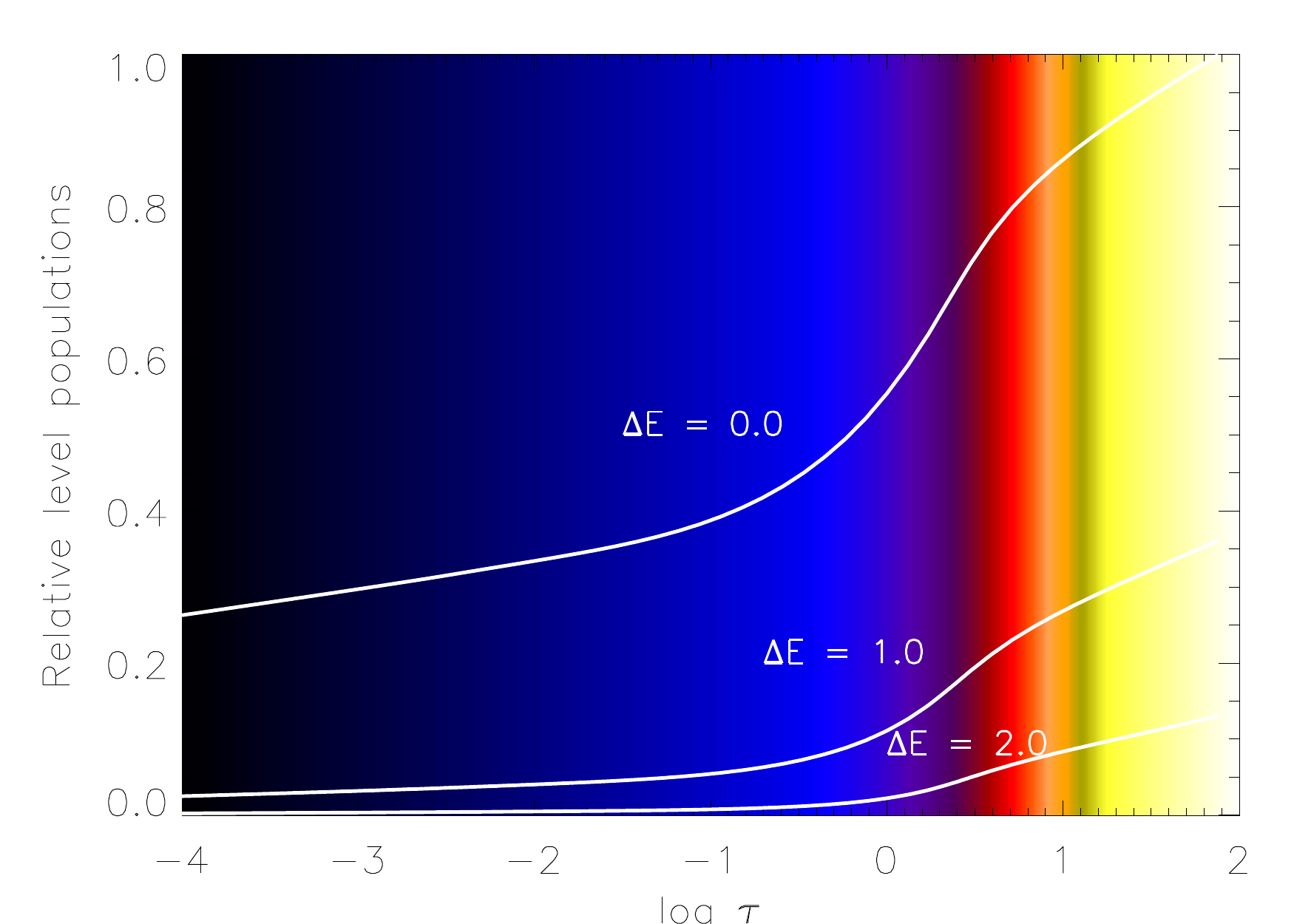}}
    \caption{Relative LTE populations for three atomic
    levels of an atom with energies $\Delta$E~=~0, 1 and 2~eV from 
    the ground state (assuming the same level degeneracy) 
    as a function of Rosseland optical depth ($\tau$) in the solar
    atmosphere. The populations are arbitrarily normalized to be unity for the ground
    level in the innermost layer shown. The color in the background indicates temperature, increasing from 
    black (about 3000~K) to white (about 10\,000~K). At the layers relevant for line formation, typically 
    $-3 < \log \tau < 0$, the populations of levels separated by 1~eV differ by a factor of about five.}
    \label{f1}
\end{figure}}

Deep enough, the atmosphere is optically thick and Local 
Thermodynamic Equilibrium (LTE) holds, so that the radiation field 
is known directly from the gas temperature. As we move upwards in the 
gravitationally stratified atmosphere, density decreases, 
photons travel longer distances 
between their emission and reabsorption or scattering, and begin 
to escape, cooling the gas (see Figure~\ref{f1}). 
Bound-bound transitions within discrete energy levels in atoms and molecules
block the light at specific wavelengths, at which the increased opacity
shifts the optical depth scale outwards, to cooler atmospheric layers,
imprinting dark stripes, absorption lines, in the spectrum, as
illustrated in Figure~\ref{f2}.

\epubtkImage{}{%
  \begin{figure}[htb]
    \centerline{\includegraphics[width=0.9\textwidth]{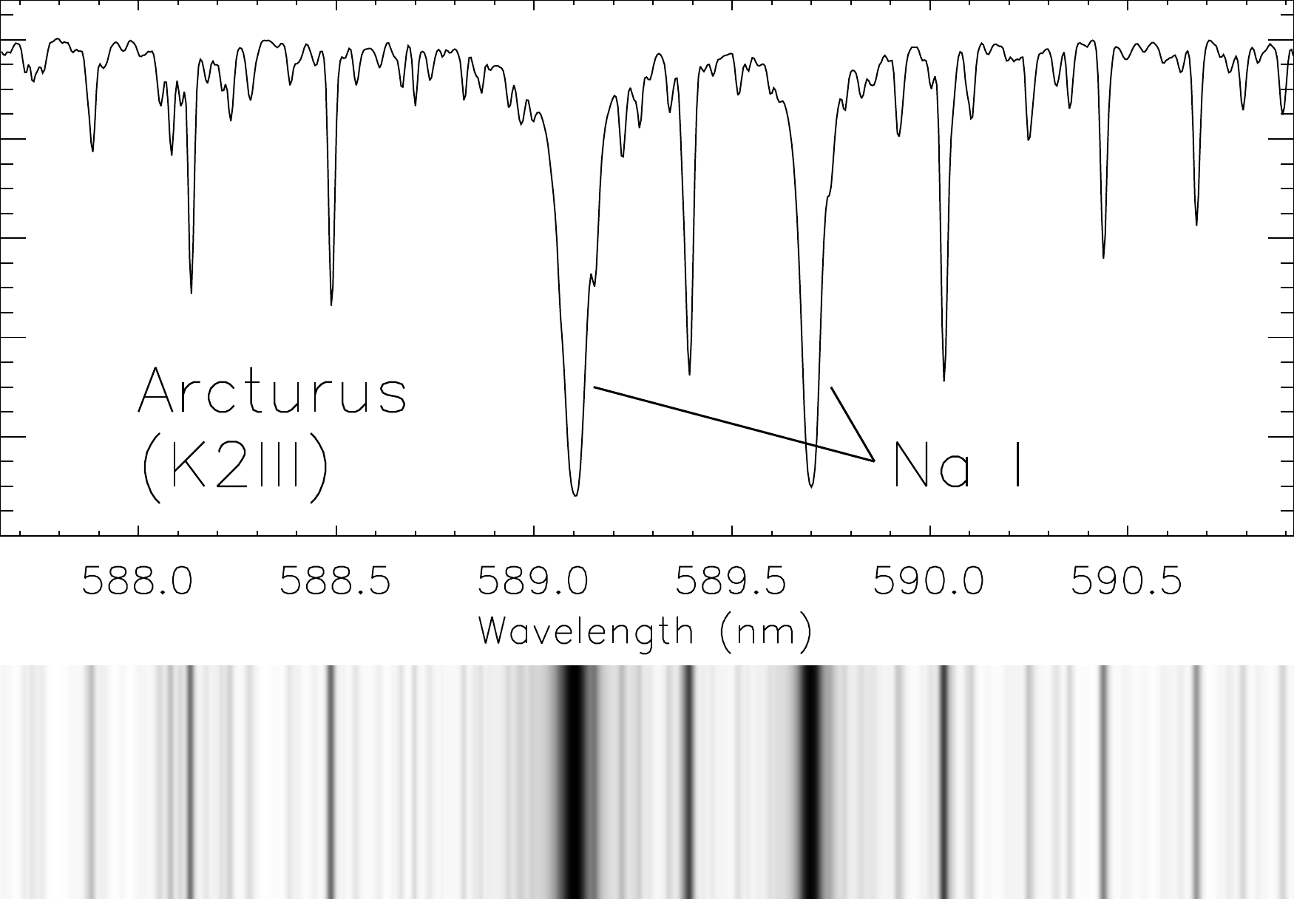}}
    \caption{A small section of a high resolution spectrum of the red
    giant Arcturus, the nearest giant at merely 11~pc from us. The strongest
    transitions in this spectral window are the Na\,{\sc i} D doublet.}
    \label{f2}
\end{figure}}

Absorption lines are easily seen at high spectral resolution, and 
constitute the main indicator of the chemical composition of the gas
in a stellar atmosphere. 
Of course, line strengths are heavily influenced by temperature, 
which dictates the relative populations of the energy levels, and 
somewhat by pressure, which has an impact on ionization 
and molecular equilibrium, and through collisional broadening. 
But for any given values of temperature and pressure, the strength of an 
unsaturated absorption line is proportional to the number of absorbers.
This simple fact, allows us to determine the chemical composition 
of stars.

There is a favorite among the stars: the Sun. This is arguably 
the best-observed and best-understood star, 
and constitutes a perfect test bed for spectral analysis techniques. 
The comparison of the Sun with other stars, and in particular, the 
derivation of accurate absolute fluxes is complicated by its 
large angular size, but otherwise it offers multiple advantages.
We can resolve its surface, determine the statistical properties 
of solar granulation, quantify limb darkening, or examine the 
center-to-limb variation of spectral lines. With the exception of
highly volatile elements (hydrogen, carbon, nitrogen, oxygen, 
and the noble gases), we can compare the derived abundances 
accessible from the solar spectrum with those in CI carbonaceous 
chondrites, which are thought to preserve the abundance 
proportions in the pre-solar nebula.

In the context of the solar system, solar photospheric 
abundances, having suffered little variation since the formation
of the pre-solar nebula,%
\epubtkFootnote{Over the solar birth, elements 
heavier than hydrogen are
expected to be reduced at the solar surface by up to 10\% 
due to diffusion \citep{2002JGRA..107.1442T}.}
serve as a reference. In a stellar context, 
the Sun is basically a calibrator for our models, both of model 
atmospheres as explained above, as well as models of stellar 
structure and evolution, which still have a number of tunable parameters.
For example, it is typically imposed that a one-solar mass model 
has the solar radius and luminosity at 4.5~Gyr, the age of the Sun.
In the context of the Galaxy, and galaxy evolution, solar abundances set 
again a reference to anchor the rest of the abundances 
-- it corresponds to the abundances in the interstellar medium 
4.5~Gyr ago at a distance from the Galactic center of about 8~kpc.

Abundances in the Sun can also be measured from emission lines
formed in the upper atmosphere, where the temperature gradient reverses
to reach millions of degrees in the solar corona, and from solar energetic
particles ejected in flares, or from particles collected from the 
solar wind. Nevertheless, these measurements are subject to larger
uncertainties than those typically involved in the analysis of
photospheric absorption lines. The additional sources of error 
are associated with our limited knowledge 
about the physical conditions in the upper solar atmosphere, required 
to interpret measurements, or with distortions introduced
before or in the process of performing the measurements. In addition, 
a poorly-understood fractionation process related to the so-called first
ionization potential (FIP) effect leads to elements with
ionization energies lower than about 10~eV to appear more abundant in solar
energetic particles and in the corona than in the underlying photosphere.

Despite that the Sun's composition sets a reference for 
abundances elsewhere in the universe, there are still important gaps 
in our knowledge of the solar composition. Solar photospheric 
abundances are usually determined relative to hydrogen, given 
that the strength of spectral lines is proportional to the ratio 
of line and continuum opacity, and H$^-$ is the dominant source 
of continuum opacity for a solar-like star. Meteoritic abundances, 
in turn, need to be referred to other elements (usually silicon), 
as meteorites have no hydrogen left. After rescaling, the 
agreement between solar photospheric abundances and CI chondrites 
abundances is better than 10\% for most elements, but 
there are notable exceptions: Sc, Co, Rb, Ag, Hf, W, and Pb 
show differences in excess of 20\%, and beyond the 
estimated uncertainties. 
In addition, there are some elements for which the photospheric
values have associated significant uncertainties 
(Cl, Rh, In, Au, and Tl), and the interesting case of Mg, 
where the estimated error bars appear slightly inconsistent 
with the observed difference between the Sun and meteorites 
\citep{2009ARA&A..47..481A,2003ApJ...591.1220L}. 

The biggest headaches are caused by the lack of information 
from CI chondrites
on the abundances of carbon and oxygen in the pre-solar nebula. 
These elements are quite abundant 
 (oxygen and carbon are the third and fourth most abundant elements 
 in the Sun, respectively, after hydrogen and helium) 
and therefore have a large weight on the overall 
\emph{metallicity} of the Sun.
A number of studies have recently reduced by 40\,--\,50\% 
the photospheric abundances for these elements, 
and solar models with updated opacities exhibit significant 
discrepancies with helioseismic determinations of the 
location of the base of the solar convection zone and the 
 helium mass fraction at the surface 
\citep[see, e.g.,][and references therein]{2009ApJ...705L.123S}.
There is indication that this problem may find a solution in the 
opacity calculations used in the interior models \citep{2015Bailey}.

In the century that humankind has been practicing spectroscopy, 
progress has been spectacular. Detectors have evolved from photographic
to photoelectric, expanding into the UV and IR wavelengths.
Telescopes and spectrographs that could only observe the Sun have been
replaced by ever larger instruments that allow observations of up to thousands
of stars simultaneously. Analyses have also evolved from using the simplest
two-layer models to sophisticated 3D (magneto) hydrodynamical simulations.
Our ability to calculate fundamental atomic and molecular constants has seen
dramatic improvements. Together, these advances open up new lines of research, and connect 
research subjects that were, just a few years earlier, seemingly unrelated.

This aims to explain the motivation for our excitement
about stellar spectroscopy, and as an overview of the procedures 
involved in the analysis of stellar spectra, in particular 
the derivation of chemical compositions of stars. Section~\ref{ingredients} 
describes the basic ingredients needed to model stellar spectra. 
Section~\ref{procedures_tools} addresses the most important 
steps that need to be followed to derive abundances, and the 
software tools available for it. The paper closes 
with an overview of ongoing observational projects, selected examples of
recent research, and a few reflections on the near- and mid-term future of 
the field.

Recent or not-so-recent but excellent 
reviews on closely related subjects have been written by 
\cite{1989ARA&A..27..701G}, \cite{1994A&ARv...6...19G}, 
\cite{2005ARA&A..43..481A}, \cite{2009ARA&A..47..481A},
\cite{2009LRSP....6....2N}, and \cite{2009LanB...4B...44L}.
A clear and concise summary of the theory 
of stellar atmospheres is given by \cite{1997LNP...497....1H}, 
and more detailed accounts are provided by 
\cite{2008oasp.book.....G} 
and \cite{2014tsa..book.....H}. 

\newpage


\section{Physics}
\label{ingredients}

The necessary physical ingredients to quantify chemical 
abundances in stars can be divided in two main categories: 
line formation and stellar atmospheres. 
Both the theory of
atmospheres and line formation are nothing but the application of
well-known physics, essentially fluid dynamics, 
statistical mechanics, and thermodynamics. Line formation and
stellar atmospheres are tightly coupled, since spectral lines
affect the atmospheric structure by blocking and redistributing radiation, 
and the atmospheric structure will largely control how lines
take shape in the spectrum of a star; they are two parts of the same
problem that we artificially split for practical purposes.

In addition to the environmental conditions of an atmosphere,
namely the stellar surface gravity and the energy per unit area 
that enters its inner boundary, modeling stellar spectra
requires knowledge of the physical constants that define
the relationship between atoms, molecules, and the radiation
field. These constants are mainly radiative transition probabilities, 
excitation, ionization and dissociation energies, as well as 
line broadening constants. 

Quantum mechanics is at the source of the microphysics data: 
distribution functions of the probabilities for atoms and molecules
to change from one state to another by collisions with other particles, 
or the absorption, emission or scattering of light by matter. 
Depending on the particular
process under consideration, these distribution functions and their
integrals are labeled collisional 
or radiative transition probabilities, line absorption profiles, or
photoionization and scattering cross-sections, but they are all
either computed from first principles, or determined
from laboratory experiments.%
\epubtkFootnote{Hybrid (semi-empirical) techniques are used sometimes.}
These input data are further discussed 
in Sections~\ref{opacity} and \ref{nlte}.

Stellar atmospheres are optically thick in their deepest 
layers, and optically thin in the outermost regions. 
Radiative transfer calculations are necessary to evaluate 
the radiation field, and determine its role in the energy 
balance, which is critical when computing model atmospheres. At that stage, 
some approximations can be adopted, and coarse frequency 
grids are commonly used. However, much more detailed calculations 
are usually necessary to compute the spectra that will be 
compared with observations, and this is commonly performed with 
dedicated codes. We deepen into this subject in Section~\ref{tools}.

\subsection{Model atmospheres}
\label{models}

In the early days, the excitation of atoms and ions was computed 
using the Boltzmann equation~(\ref{nage}) for a single value of temperature 
\citep[see, e.g.,][]{1929ApJ....70...11R}.
Based on theoretical developments by Schwarzschild, Milne, and Eddington, 
\cite{1931MNRAS..91..836M} 
started the construction of non-gray model atmospheres, 
which have grown in sophistication ever since. Line-blanketed models, 
those taking into account the effect of line opacity on the atmospheric structure, 
made an appearance in the 1970s 
\citep{1969tons.conf..377C,1969ApJS...18..127P,
1972ApJ...176..629A,1970mase.book.....P,
1975A&A....42..407G,1979ApJS...40....1K}
and continued to evolve, hand in hand with more complete sets of opacities and 
improvements in computational facilities providing a 
better description of the radiation field
\citep[see, e.g.,][but we refer to Hubeny and Mihalas' book for a more
exhaustive list]{1992IAUS..149..225K,1999ApJ...512..377H,
1999ApJ...525..871H, 
2004astro.ph..5087C,2008A&A...486..951G}.

As mentioned in the introduction, the 
fundamental parameters than define a model atmosphere are
the energy flux that traverses it (parameterized as a function of the 
effective temperature $T_{\mathrm{eff}}$), 
 surface gravity, 
and chemical composition. Only a few elements in 
the periodic table are relevant, and therefore this latter parameter
is usually simplified to just one quantity, the 
metallicity:%
\epubtkFootnote{Occasionally the metal mass fraction is used: 
$Z=\sum_i A_i N_i/N_{\mathrm{H}}$, where $A$ is the atomic mass in units of
the hydrogen mass}
\begin{equation}
[\mathrm{Fe/H}] = \log \left(N_{\mathrm{Fe}}/N_{\mathrm{H}}\right) 
  - \log \left(N_{\mathrm{Fe}}/N_{\mathrm{H}}\right)_{\odot}.
\end{equation}
where $N_X$ represents number density of nuclei of the elements X. 
This is practical because reasonably constant ratios are found 
between the abundances of iron and most other metals, 
and iron is in general the element easiest to
measure with precision due to the large number of iron lines 
visible in stellar spectra.

Models for hot stars, for which it is quite important to consider 
departures from LTE, have also made parallel progress since the
early developments \citep[see, e.g.,][]{1969ApJ...158..641A}, 
to more recent fully-blanketed calculations 
\citep{1995ApJ...439..875H,1995ApJ...439..905L,1997ApJ...485..843L,
2003ApJS..146..417L,2007ApJS..169...83L}.
At the hot end of the spectrum, it becomes 
necessary to account for stellar winds,
and unified models have been presented by, e.g., 
\cite{1986A&A...164...86P}, \cite{1989A&A...226..162G}, \cite{1996A&A...305..171P}, 
or 
\cite{2012IAUS..282..229H}. 
LTE models for white dwarfs, where the high density helps to maintain
LTE valid, have also been calculated 
\citep[see][and references therein]{2010MmSAI..81..921K}.
A general code, with an emphasis on parallel computations,
has been described by \cite{1997ApJ...483..390H} and \cite{2001ApJS..134..323H}.
On the cool end of the mass range, models for low-mass stars, 
brown dwarfs and planets 
have seen extraordinary development in the last decade
\citep[e.g.,][]{1997ARA&A..35..137A,2015ARA&A..53..279M}.

In parallel with refinements in 1D models, 2D and 3D models have emerged 
and started to be applied to stars 
\citep{1990A&A...228..155N,1990A&A...228..184D,1990A&A...228..203D,
1989ApJ...342L..95S,1998ApJ...499..914S,2000A&A...359..743A, 
2013A&A...557A..26M, 
2004A&A...414.1121W,2012JCoPh.231..919F,2013A&A...557A...7T}.
These models deal with the hydrodynamics 
of stellar envelopes, and describe convection, which accounts 
for up to $\sim$~10\% of the energy transported outward in 
the solar photosphere, from first principles.
The onset of convection is caused by the recombination 
of hydrogen at the solar surface. 
When hydrogen becomes ionized, the number
of free electrons increases dramatically, and so does the blocking effect on
radiation caused by electron scattering.
This takes place when the gas temperature is about 10\,000~K, as one would expect
from Saha's equation~(\ref{saha}), and 100~km above those layers 
the average temperature in the solar atmosphere drops to nearly 5000~K.

The low viscosity of the stellar plasma, and consequently high Reynolds numbers,
is responsible for the turbulent behavior of the gas in the solar
envelope, including deep photospheric layers. Compared to
the classical 1D (hydrostatic equilibrium) models, 3D simulations fully
account for the effect of velocity fields on the atmospheric structure
and spectral lines, which get broadened and (typically) blue-shifted. 
This is illustrated in Figure~\ref{blueshifts}, where
spectra computed for a solar-like star using a 1D classical model
and a \textsc{CO5BOLD} simulation of surface convection are compared. 
On the other hand, the fact that the model is
3D (spatially) and time-dependent, necessarily requires simplifying 
the calculation of the radiation field to evaluate the energy balance, 
and the radiative transfer
is typically solved for only a few (wisely chosen) frequency bins.

\epubtkImage{}{%
 \begin{figure}[htb]
    \centerline{\includegraphics[width=0.9\textwidth]{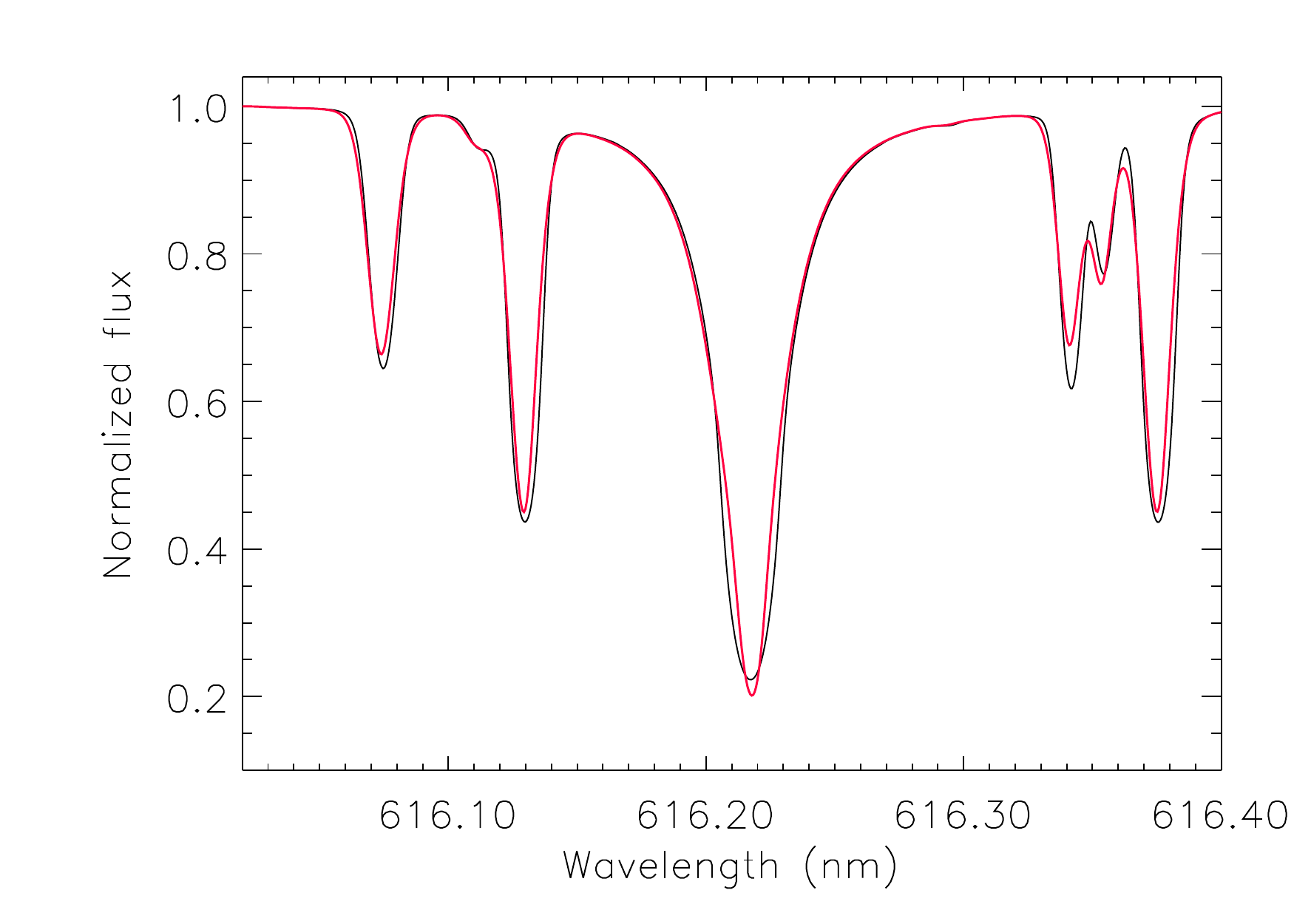}}
    \caption{
Spectra in the vicinity of the Ca\,{\sc i}~$\lambda$616.2~nm line computed for a solar-like 
star using a classical hydrostatic model (black) and a \textsc{CO5BOLD} hydrodynamical 
simulation \citep[red; see][]{2013A&A...557A...7T}.}
    \label{blueshifts}
\end{figure}}

In terms of availability, not all models are made equal. The most 
complex and costly-to-calculate models tend to be custom-made. On the other hand, 
 large grids of 1D models are publicly available from
Kurucz%
\epubtkFootnote{\url{kurucz.harvard.edu} and
  \url{www.iac.es/proyecto/ATLAS-APOGEE/}, 
but see also \url{www.univie.ac.at/nemo/}} 
($3500\le T_{\mathrm{eff}}\le 50\,000\mathrm{\ K}$), 
the Uppsala group \epubtkFootnote{\url{marcs.astro.uu.se}}, and 
Hubeny \& Lanz\epubtkFootnote{\url{nova.astro.umd.edu}} 
(NLTE models; $15000\le T_{\mathrm{eff}}\le 55\,000\mathrm{\ K}$). 
Others are available as well, but have been less frequently used, e.g., 
the PHOENIX models published by \cite{2013A&A...553A...6H} or those at 
France Allard's web site.%
\epubtkFootnote{\url{http://perso.ens-lyon.fr/france.allard/}}
Many other families
of models, in particular 3D models, are proprietary, but some can be 
obtained directly from their authors upon request or negotiation. 
The same is true for the model-atmosphere codes. Kurucz's codes
and Tlusty are publicly available from the corresponding web sites%
\epubtkFootnote{Note that Kurucz's codes have now been ported to
  Linux, and there is documentation available. See
  Section~\ref{abundances}}.

\subsection{Line formation theory}
\label{radiative_transfer}

Radiation plays a key role in determining the atmospheric structure,
and therefore has to be described adequately in calculations 
of model atmospheres. Conversely, once we have a model structure it
becomes feasible to perform much more detailed 
radiation transfer calculations to predict stellar spectra. 
Armed with an appropriate model atmosphere and the necessary line 
data, we can proceed to compute spectra that can be 
compared with observations to infer the chemical makeup of stars.

The detailed shapes and strengths of spectral lines depend on the
thermodynamic structure of the atmosphere, which under LTE 
 will dictate the ionization and excitation of atoms
and molecules, as well as thermal and collisional broadening. As
opposed to the situation with radiative line transition probabilities, 
laboratory experiments on line broadening by hydrogen atoms, the main perturber
in solar-like stars, are extremely hard, and the 
sensitivity of lines to collisions is usually derived from calculations, 
as illustrated, e.g., by \citet{1998MNRAS.296.1057B} or \citet{2005A&A...435..373B}.
As mentioned in Section \ref{models}, convective motions in late-type stars 
will also directly broaden 
the line profiles, and the lack of them in 1D models is compensated
by introducing two \emph{ad hoc} parameters: micro- and macro-
turbulence.

Depending on the geometry of the model, the radiative transfer
equation for quasi-static conditions
\begin{equation}
  \frac{\partial I_{\nu}}{\partial x} = \eta_{\nu} - \kappa_{\nu} I_{\nu} \,,
\end{equation}
where $I_{\nu}$ is the intensity of the radiation 
field (energy per unit area, per unit of time, per unit of frequency,
per unit of solid angle) at a frequency $\nu$, 
$x$ is the spatial coordinate along the ray, 
$\kappa_{\nu}$ is the opacity, 
and $\eta_{\nu}$ the emissivity at that frequency $\nu$,
will need to be solved somewhere between three (three
ray inclination angles for a plane-parallel model neglecting scattering) 
and millions (3D time-dependent models) of times per frequency. 
Departures from LTE, considered in Section~\ref{nlte}, usually involve 
many more calculations, as the statistical equilibrium 
equations are solved iteratively, requiring multiple 
evaluations of the radiation field at all depths.
 
A number of LTE codes for solving the radiation transfer equation and 
computing detailed spectra are available. Typically, there is one such 
code associated
with each of the packages used to compute model atmospheres. For example, 
there is \textsc{Turbospectrum} for
\textsc{MARCS} (Uppsala) models, \textsc{SYNTHE} for Kurucz models, 
and \textsc{Synspec} for Tlusty models.

\subsubsection{Opacity: lines and continua}
\label{opacity}

Under the assumption of LTE, two thermodynamic variables (e.g., temperature 
and gas pressure) determine the ionization and excitation fractions, which, 
together with the line collisional broadening, 
can be used to calculate how opaque 
to radiation is matter. Opacity is characterized by the likelihood that 
a photon is absorbed per flying unit length, which is of course a 
function of frequency: $\kappa_{\nu}$. Restricting our discussion to LTE, 
the thermal component of the emissivity, $\eta_{\nu}$, is trivially calculated from the opacity,
as the source function, 
$S_{\nu}$, is equal to the Planck's law, and therefore only depends on temperature
\begin{equation}
S_{\nu} \equiv \eta_{\nu}/\kappa_{\nu} = B_{\nu}(\mathrm{T}) 
= \frac{2h\nu^3}{c^2} \frac{1}{e^{\frac{h\nu}{k\mathrm{T}}} -1} \,,
\end{equation}
where T is temperature, $\nu$ is frequency, 
$c$ is the speed of light, $h$ is the Planck constant, 
and $k$ is the Boltzmann constant.

All absorption is divided into transitions between bound energy levels 
(\emph{lines}), or between bound levels and the continuum 
(\emph{continua}). Transitions between short-lived unbound 
states, i.e., the absorption of a photon by an atom, 
ion, or molecule that temporarily forms by the proximity of two 
particles (e.g., an atom and an electron in the case of H$^{-}$), 
cannot only happen, but may contribute significant \emph{free-free}
opacity. The strength of a spectral line is critically dependent on the 
line-to-continuum opacity, and the line opacity can be written as
$l_\nu = N \alpha$, where $N$ is the number density of absorbers
(i.e., the abundance of ions or molecules with the appropriate excitation 
ready to absorb the photon), and $\alpha$ is the absorption 
coefficient per absorber (i.e., the cross-section)
\begin{equation}
\alpha (\nu) = \frac{\pi e^2}{m c} f \phi({\nu}) \,,
\label{alpha}
\end{equation}
where $e$ and $m$ are the charge and mass of the electron, 
$f$ is the oscillator strength or $f$-value (proportional to the 
radiative transition probability $A$), and $\phi({\nu})$ is the Voigt 
function, the result of the convolution of a Gaussian 
(due to thermal broadening) and a Lorentzian (due to natural broadening and 
collisional damping), describing the shape of the absorption 
probability distribution around the central frequency of the line.

Under LTE, the number density of the absorbers can be split 
into three factors
\begin{equation}
N = \frac{g}{u_j} N_j e^{-\frac{E}{kT}} \,,
\label{nage}
\end{equation}
where $j$ is the charge (in units of the charge of the electron) 
of the ion (i.e., $j=0$ for neutral atoms), 
$N_j$ is the abundance (number density) and 
$u_j$ the partition function for the ion $j$, 
$E$ is the energy of the state (above the ground level), and $g$ is its degeneracy.
The fraction of nuclei of a given element that are tied in an ion $j$ 
is given by Saha's equation \citep[see][]{2012arXiv1209.1111F}
\begin{equation}
\label{saha}
\frac{N_j}{\sum_i N_i} = 
\frac{\gamma^j u_j}{\sum_i \gamma^i u_i e^{-\frac{\beta_i}{kT}}} 
e^{-\frac{\beta_j}{kT}} \,,
\end{equation}
where $\gamma = 2/(n_e h^3) (2\pi m k)^{\frac{3}{2}} 
T^{\frac{3}{2}}$,
$n_e$ is the number density of electrons, 
and $\beta_j = \sum_{i=0}^j \chi_i$, where $\chi_i$ is the 
energy required to rip an electron from the ion $i-1$ ($\chi_0= 0$) in 
the ground state.

Radiative transition probabilities, or $f$-values, have a direct influence 
on line strength, just the same as the number density of absorbers. 
Unfortunately, only 
for simple (light) elements it is feasible to calculate 
accurately, using quantum mechanics, radiative transition probabilities. 
For more complex atoms, most data need to be determined from laboratory 
experiments, a difficult and time-consuming work.
Traditionally, the US National Institute for Standards and Technology (NIST) 
has done useful compilations, formerly available through a series of books, 
and now online.%
\epubtkFootnote{\url{www.nist.gov}}
Kurucz (and colleagues) have made a 
tremendous effort to augment significantly the number of lines 
with available transition probabilities, by performing semi-empirical 
calculations \citep[see, e.g.,][]{1975SAOSR.362.....K}, 
even though those are of significantly lower quality than laboratory
measurements. Another useful resource for transition probabilities is the Vienna
Atomic Line Database \citep[and references therein]{2008JPhCS.130a2011H, 
2011KIzKU.153...61R}, 
which tries to catch up with the 
many papers in the literature on line data, whatever its source.

Calculations of bound-free absorption (photoionization) made for light 
elements ($Z \le 26$) in the context of the Opacity Project 
\citep[OP, see][and references provided there]{2005MNRAS.362L...1S,2005MNRAS.360..458B}, 
have been available for a long time \citep{2001APS..DMP.S5079M}. 
This work has been extended more recently by the Iron Project team 
and Sultana Nahar maintains an online 
database%
\epubtkFootnote{\url{http://www.astronomy.ohio-state.edu/~nahar/nahar_radiativeatomicdata/}}
\citep{2011CaJPh..89..439N}. 
The accuracy of the energies computed by these \textit{ab initio} calculations 
is about 1\%, and therefore it has been recommended that the 
photoionization
cross-section be smoothed accordingly \citep[see, e.g.,][]{1998ApJS..118..259B}.
\cite{2003ApJS..147..363A} have applied this procedure to the OP data,
and compiled model ions for use with the codes 
\textsc{Synspec} and \textsc{Tlusty} \citep{1995ApJ...439..875H}. 
Free-free absorption, as mentioned above, can be important for 
abundant species, such as H, H$^-$, or He$^-$.
In fact, H$^{-}$ bound-free absorption is the dominant source of continuum opacity in the optical 
for the solar photosphere, followed by atomic H, and free-free opacity from the same
ion dominates in the IR. Figure~\ref{f3} 
illustrates the estimated impact of bound-free 
absorption by different atoms relative to the total hydrogen 
(atomic H and H$^{-}$) continuum opacity on the solar spectrum.

\epubtkImage{}{%
  \begin{figure}[htb]
    \centerline{\includegraphics[angle=90,width=0.9\textwidth]{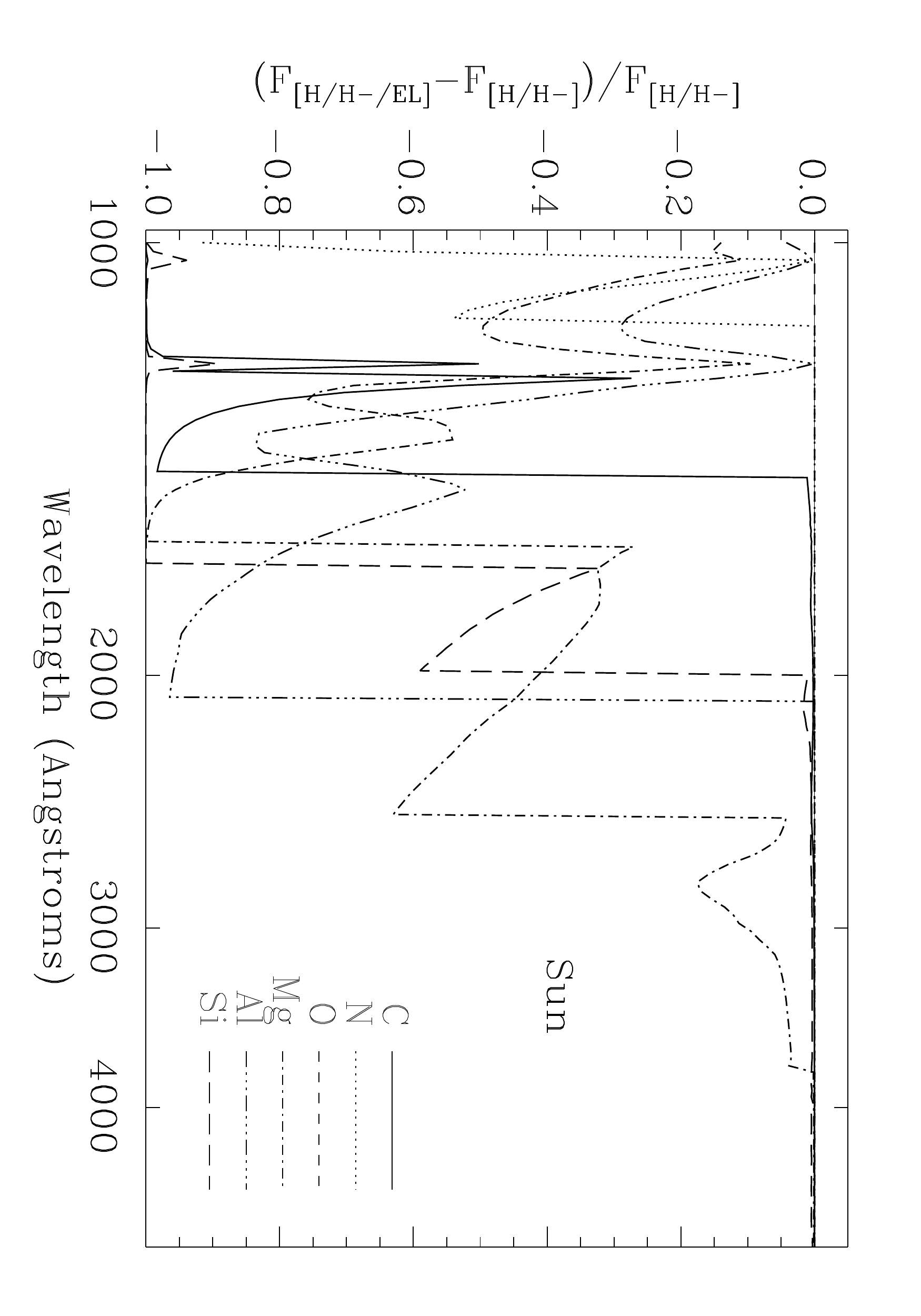}}
    \caption{Impact of continuum opacity of different atoms on the 
    shape of the solar flux relative to the dominant source of
	continuum opacity, bound-free H and H$^{-}$ absorption.
	Image reproduced with permission from \cite{2003ApJS..147..363A}, copyright by AAS.} 
    \label{f3}
\end{figure}}

Although we are discussing mainly \emph{upward} transitions
because those are directly observed in the photospheric spectrum, 
the opposite \emph{downward} processes are also taking place
-- at about the same rate, to have equilibrium. 
Along with radiative excitation and ionization, we have 
radiative de-excitation and recombination.
Spontaneous (\emph{downward})
radiative transitions also occur. Finally, collisional processes
are always competing, though we do not need to take them into account
to compute the opacity under LTE, as 
each process is exactly balanced by the opposite, i.e., \emph{detailed
balance} holds (the sum of all radiative transitions between a given
energy state and all others must be equal to the sum of radiative
transitions from all other states to that one, independently from the 
collisional transitions).


\subsubsection{Departures from LTE}
\label{nlte}

Under the assumption of LTE, radiation and matter are coupled locally, and 
the source function is equal to the Planck function.
But no physical law enforces LTE, in particular when the photon's mean free path
is long. From a more general perspective, we can consider a case in which
the level populations are time independent, i.e.,
the sum of all radiative and collisional transitions from an energy level and
all others equals the sum of all transitions from all other levels to that one
\begin{equation}
n_i \sum_{j} (R_{ij} + C_{ij}) = \sum_{j} n_j (R_{ji} + C_{ji}) \,,
\label{rate_equations}
\end{equation}
where $R_{ij}$ and $C_{ij}$ indicate the radiative and collisional rates, respectively, 
between the states labeled $i$ and $j$. Note that $R_{ij}$ is closely related
to the line absorption coefficient introduced in Eq.~(\ref{alpha}); 
it is proportional to the 
product of $\alpha$ times the photon density, integrated over the line profile
(plus a contribution from spontaneous emission if $i>j$).
The system of equations~\ref{rate_equations} for $n$ states 
includes $n-1$ linearly dependent equations, and therefore it needs to
be closed by adding a total particle number conservation equation.

Solving the statistical equilibrium equations requires many data
which are not needed under LTE. To evaluate all possible 
transition rates across states and ions we need not only radiative
cross-sections (i.e., transition probabilities and photoionization cross sections) 
but also collisional strengths (collisional transition probabilities, if you will). 
Given their high density, electrons and hydrogen 
atoms are the main particles responsible for inelastic collisions.
As discussed in Section~\ref{radiative_transfer} for line broadening 
(elastic collisions), laboratory experiments on inelastic collisions 
with atomic hydrogen are very hard to carry. Approximate formulae
exist for electron-impact collisional excitation and ionization 
 \citep[see][]{2003ApJS..147..363A}, although these are, at best, only good
to provide order-of-magnitude estimates. 
Quantum mechanical calculations of atomic structure can provide 
much better estimates for collisional strengths 
for electron impacts 
\citep[see, e.g.,][]{2003ApJS..148..575Z,2007A&A...462..781B,2014A&A...572A.103B,2015A&A...579A..53O}, but more involved 
methods, requiring an adequate knowledge of 
the relevant molecular potentials, are needed to derive reliable
estimates for hydrogen collisions \citep[e.g.,][]{2006ApJ...647.1531K,2012A&A...541A..80B}.

Under solar photospheric conditions, hydrogen atoms dominate by number over free electrons,
but inelastic hydrogen collisions tend to have a limited role
on line formation \citep{2004A&A...423.1109A,2013A&A...554A.118P}.
Figure~\ref{clv} illustrates, however, that their effect is visible
in the changes of strength experienced by lines such as the O\,{\sc i} triplet
at 777~nm as a function of heliocentric angle. 
At lower metallicities, higher pressure and a lower density of free
electrons combine to make the role of hydrogen collisions much 
more important, as discussed for example by \cite{1999A&A...350..955G}.
\cite{2005ApJ...618..939S}
predict abundance corrections due to NLTE effects on Fe\,{\sc i} in the metal-poor
subgiant HD~140283 as large
as 0.6~dex for a 1D model atmosphere, and as large as 0.9~dex for
a 3D model, in the absence of inelastic hydrogen collisions. 

\epubtkImage{}{
  \begin{figure}[htb]
    \centerline{
      \includegraphics[angle=90,width=0.5\textwidth]{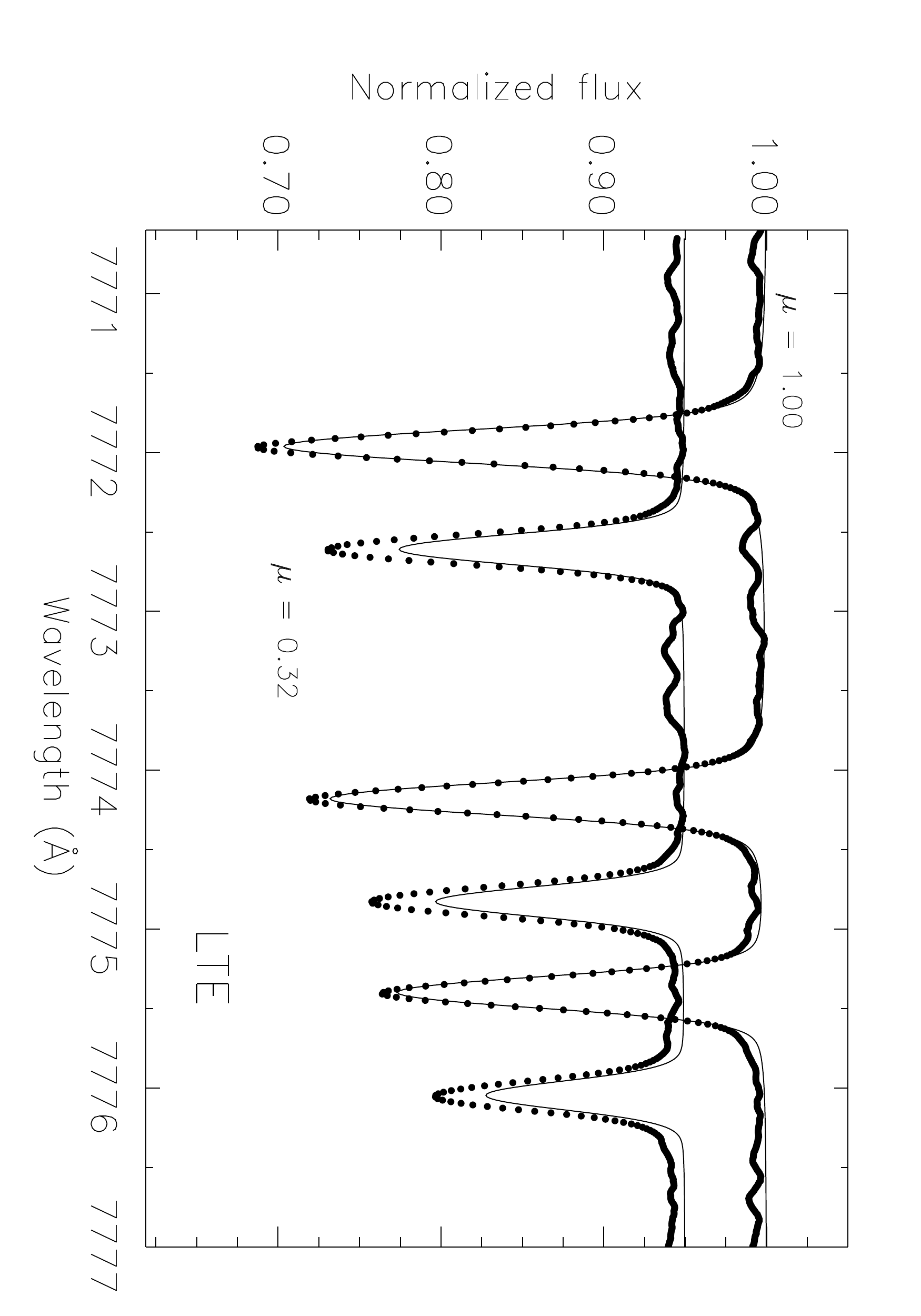}
      \includegraphics[angle=90,width=0.5\textwidth]{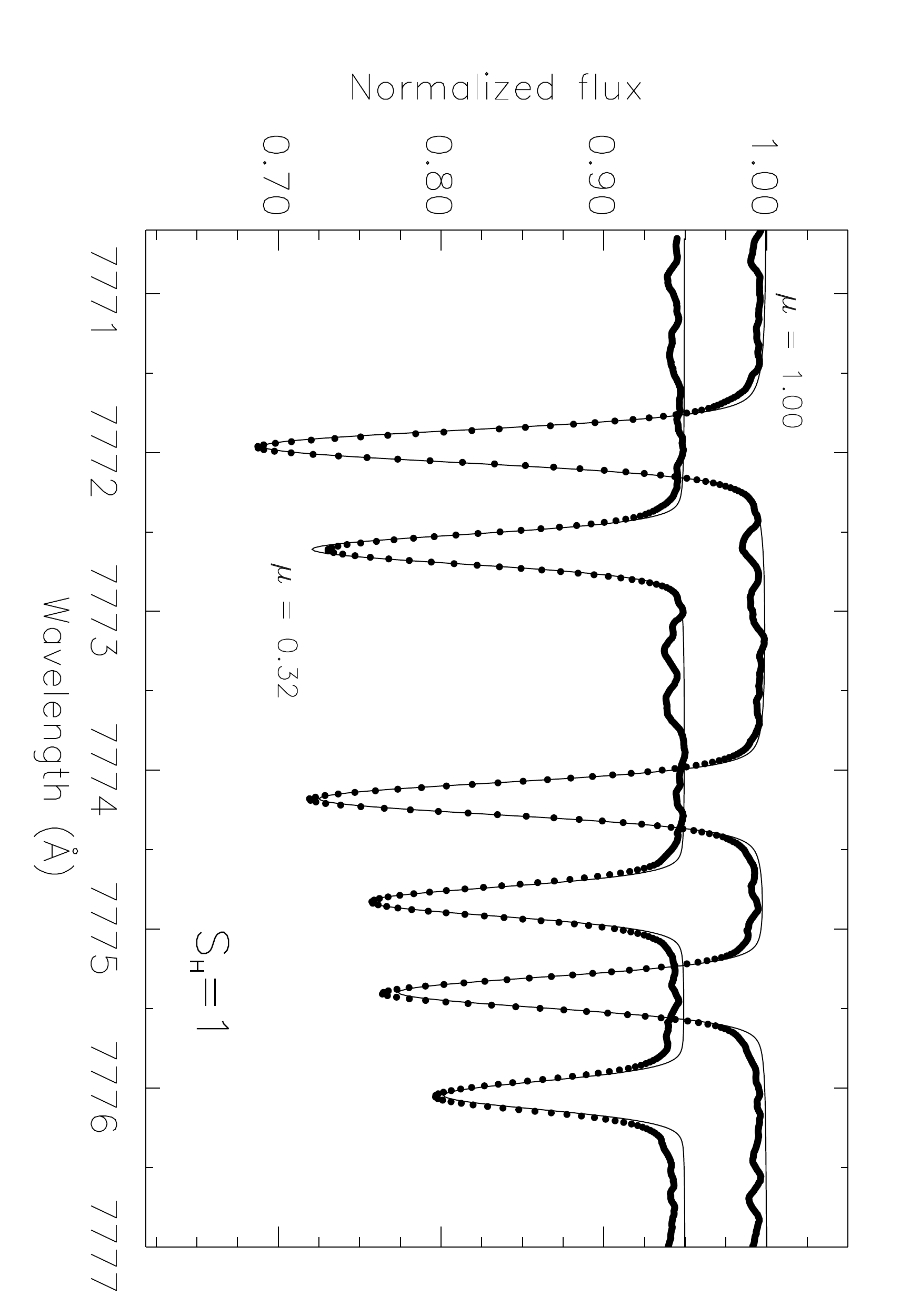}
    }
    \caption{Observed behavior of the the O\,{\sc i} triplet at 777~nm in the Sun as a function of 
	the inclination of rays from the solar surface ($\mu= \cos \theta$,
 where $\theta$ is the angle between the observer and the
normal to the solar surface). The filled circles correspond to observations, and the 
solid lines to models. The observations at disk center ($\mu=1$) are reproduced in all cases, 
by changing the oxygen abundance, and the comparison near the limb ($\mu=0.32$) tests
the realism of the calculations. The behavior of the lines can be reproduced considering 
inelastic collisions with hydrogen atoms in the rate equations. The factor $\mathrm{S_{H}}$
is introduced to enhance the approximate collisional rates from a Drawin-like formula as 
proposed by \cite{1984A&A...130..319S}. 
Image reproduced with permission from \cite{2004A&A...423.1109A}, copyright by ESO.} 
    \label{clv}
\end{figure}}

Departures from LTE can be very important in some situations. If they
involve important species (say, hydrogen, helium, iron), their impact can be
profound and the statistical equilibrium equations may need to be
solved together with the model atmosphere problem, as it is
usually done for hot stars. In many other cases,
the departures affect trace species and particular transitions of interest, 
which may induce a modest
effect on the overall atmospheric structure, but anyway must be considered
in order to avoid large systematic errors in the calculated line profiles, 
and therefore the derived abundances. In this latter case, one can adopt
the thermal structure of the atmosphere plus a second thermodynamic 
variable (usually the electron density) and solve the statistical 
equilibrium equations~(\ref{rate_equations}) for the relevant ions 
in isolation from the model atmosphere, what is usually referred to
as the \emph{restricted} NLTE problem.

There are NLTE codes available for solving the \emph{full} NLTE problem 
(statistical equilibrium plus atmospheric structure) and some
limited to the \emph{restricted} 
NLTE problem. Among the former, we find the publicly available codes
\textsc{CMFGEN}%
\epubtkFootnote{\url{http://kookaburra.phyast.pitt.edu/hillier/web/CMFGEN.htm}}
\citep{2012IAUS..282..229H}
the suite of Munich codes%
\epubtkFootnote{\url{http://www.usm.uni-muenchen.de/people/adi/Programs/Programs.html}}
\citep{1998ASPC..131..258P}, 
and Tlusty%
\epubtkFootnote{\url{http://nova.astro.umd.edu/}}
\citep{1995ApJ...439..875H}, 
 three packages originally developed in the context of hot stars. 
The code \textsc{MULTI}%
\epubtkFootnote{\url{http://folk.uio.no/matsc/mul22/}}
\citep{1992ASPC...26..499C} 
deals with the \emph{restricted} NLTE problem, and has been widely
used in the context of late-type atmospheres.

\newpage


\section{Working Procedures}
\label{procedures_tools}

With the basic blocks described above -- adequate opacities, 
and codes to compute model atmospheres and spectra -- the only
remaining input is an observed spectrum, and a work plan.

The various elements entering in the analysis are tightly
coupled. To calculate a model atmosphere, we need to
know the basic atmospheric parameters and abundances,
precisely the very variables we seek to derive from the
spectral analysis. Thus, it is only natural to address the problem
 iteratively. First, atmospheric parameters
are guessed, then a spectrum computed and confronted with the
observations to determine the parameters, repeating the cycle
as needed. Such process can, in some situations,
be simplified, provided we are confident that some parameters 
are decoupled from others -- for example, when we derive the
abundance of iron in a solar-like star from lines of singly-ionized
Fe ions, which have only a weak sensitivity to the surface gravity.
In some cases, parameters derived from external data, free from 
systematic errors that model spectra are subject to, are 
preferred, and that also simplifies the analysis, eliminating the
need for iteration.

The amount of information available directly from a spectrum
will be limited by its spectral coverage, signal-to-noise, 
and resolution.%
\epubtkFootnote{One may define a \emph{power-to-resolve} $P$ that combines 
these three parameters to quantify approximately the information content in 
a spectrum \citep{HERAEOUS2015}.}
In addition, 
spectra show a strong response to some parameters 
(e.g., the effective temperature),
and a much weaker or null reaction to others (e.g., the
abundance of ytterbium). Tracking correlations is extremely
important to arrive at reliable uncertainty estimates in the 
inferred abundances. 

Because of the low temperature and shallow temperature structure in the outer
layers of the photosphere, lines with fairly opaque cores saturate,
i.e., their strength does not grow linearly with the abundance of
the absorbers, and therefore relatively weak lines are preferred.
Weak lines demand high spectral dispersion, which usually 
implies narrow spectrograph slits and high frequency variations 
in the instrument's throughput 
(i.e., poor accuracy in flux calibration), and more limited 
spectral windows. For this reason, the analysis of 
high-resolution spectra is usually supplemented with
lower-dispersion spectrophotometry or photometry.

In what follows, we will briefly split the abundance
determination process into three steps: gathering data, 
choosing atmospheric parameters, and deriving abundances.

\subsection{Obtaining stellar spectra and reference data}
\label{observing}

Because the atmospheric parameters are related to the fundamental 
stellar parameters, and these can be constrained from
many other data independent from spectra, we should
compile all the relevant information available from the literature. 
If we aspire to provide a physically consistent picture of the observed
objects, the atmospheric parameters used in the spectroscopic
analysis need to be consistent both with the spectrum and with
the available external data, whenever possible.

Examples of useful data are: trigonometric parallaxes from astrometric
measurements, angular diameters from interferometry, or mean densities
from asteroseismology \citep[see, e.g.,][]{2010MNRAS.405.1907B,2014ApJS..215...19P,
2015MNRAS.451.3011C,2015MNRAS.452.2127S}.
However, be aware that in most cases deriving
fundamental quantities such as radii, masses, or distances 
from these observations involves to some extent the use of 
models of stellar structure
and evolution, subject to their own issues and systematic errors.
If the stars of interest are members of binary systems, there may
be very useful constraints on the stellar masses from the system
dynamics, and if we are lucky 
enough to have an eclipsing system, then masses and radii may be
constrained with exquisite accuracy
\citep{1980ARA&A..18..115P,1991A&ARv...3...91A,2010A&ARv..18...67T}, 
but the likelihood for such alignment is very low.

Digital data can be handled and stored with ease, and as
instruments become more efficient and observations more
homogeneous, there are more and more data archives that offer
collections of stellar photometry and spectra of different kinds.
In Section~\ref{libraries} we provide a list of some of the
most popular collections of spectra for nearby and bright stars.

Photometric information, and most of all, spectrophotometric information,
is of high value when it comes to constraining atmospheric parameters,
in particular effective temperature, although the interstellar 
extinction complicates the analysis. 
There are large libraries with spectrophotometry for stars, some
of which are mentioned in Section~\ref{libraries}.

A word should be said here about photometric calibrations for
deriving stellar effective temperatures from photometry.
Direct calibrations are rarely used, due to the high sensitivity 
of the model atmospheres and irradiance calculations to 
the input physics, and in particular the equation of state. 
The most widely-used calibrations are based on the so-called
infrared flux method 
\citep{1980A&A....82..249B,1999A&AS..140..261A, 
2005ApJ...626..465R,2009A&A...497..497G,2010A&A...512A..54C}, 
which exploits a more robust prediction from the models, the
ratio of the monochromatic flux at a given infrared wavelength 
to the bolometric flux of a star.

In many cases, spectra for the stars of interest are not available from
existing data bases, and we will need to obtain new observations.
Observing facilities are continuously progressing. However, 
spectral resolution is not increasing as spectrographs 
become larger to mate with larger telescopes. For a given grating, 
the product of the
resolving power and the slit width are proportional to the ratio of 
the collimator and telescope diameters, making it difficult
to reach high dispersion without using narrow slits and losing light. 
In addition, there 
has only been modest progress in feeding multiple objects
to high-resolution spectrographs, at least in comparison 
with the massive multiplexing capabilities now available 
and planned for lower dispersion instruments.%
\epubtkFootnote{MUSE for the VLT employs 
image slicers to feed 24 spectrographs at once \citep{2012Msngr.147....4B}, 
and VIRUS for the Hobby-Eberly telescope has 
some 30\,000 fibers feeding 75 spectrographs simultaneously \citep{2012AAS...21942401H}.}
Many observatories offer high-resolution spectrographs, but 
reviewing the existing choices is out of the scope of this paper. 
We will go back to discussing some of the largest projects, ongoing and planned, in Section~\ref{data}.

\subsection{Atmospheric parameters}
\label{params}

Provided with a wide spectral range, it may be feasible 
and convenient to derive all the relevant atmospheric
parameters needed to calculate a model atmosphere from 
the very same high-resolution spectra that will be 
later used to derive abundances. Nevertheless, alternative
paths are useful for two reasons: as a sanity
check (remember that the analysis of the high-resolution spectra
will have associated caveats, as it involves many approximations), or
simply because a high-resolution spectrum will have a limited
sensitivity to some parameters, with surface gravity being the
most typical case. 
We mentioned in the previous section several possibilities 
to constrain the atmospheric parameters from measurements
other than high-resolution spectra. Here we will highlight
the most useful features available in spectra when 
lines are resolved.


The effective temperature is usually derived from the 
excitation balance of atomic iron, calcium, silicon or titanium,
i.e., by requiring that lines with different excitation energy
give the same abundance.
This technique simply takes advantage 
of the Boltzmann equation [see Eq.~(\ref{nage})], where T is 
the temperature in the relevant
atmospheric layers for the lines, which is of course tightly
coupled to the effective temperature of the star $T_{\mathrm{eff}}$.
Unfortunately, the de-saturation of strong lines induced 
by small-scale motions of the iron atoms (micro-turbulence), 
complicates somewhat the performance of this method, which is sensitive
to this parameter, as well as to departures from LTE.

The damping wings of hydrogen lines, 
largely controlled by Stark broadening induced by elastic collisions 
with free electrons, but
also affected by collisions with the surrounding hydrogen atoms, 
are also highly dependent on the thermal structure in layers
close to Rosseland optical depth $\tau \sim 1$, which provides an excellent thermometer
\citep[see][and references therein]{1993A&A...271..451F,1994A&A...285..585F, 
2000A&A...363.1091B,2002A&A...385..951B, 
1983A&A...127..263S,1994A&AS..104..509S}. 
This is illustrated in Figure~\ref{balmer}.
Here a difficulty for one-dimensional models, in addition to 
getting right the microphysics for Stark broadening, is to account
properly for the effect of convective motions on 
the thermal structure of deep photospheric layers.
Three-dimensional models based on hydrodynamics get rid 
of this problem. Investigations of departures from
LTE on the wings of Balmer lines have also shown that
these effects need to be considered if high accuracy is sought 
\citep[see][]{2004ApJ...609.1181P,2007A&A...466..327B}. 
In any case, \citet{2011A&A...531A..83C} have demonstrated that the scale can be
calibrated using external data.
Most sensitive, but hard to calibrate in absolute terms,
is perhaps the technique based on line-depth ratios 
\citep[see, e.g.,][]{2001PASP..113..723G}.

\epubtkImage{}{
  \begin{figure}[htb]
    \centerline{\includegraphics[angle=90,width=0.9\textwidth]{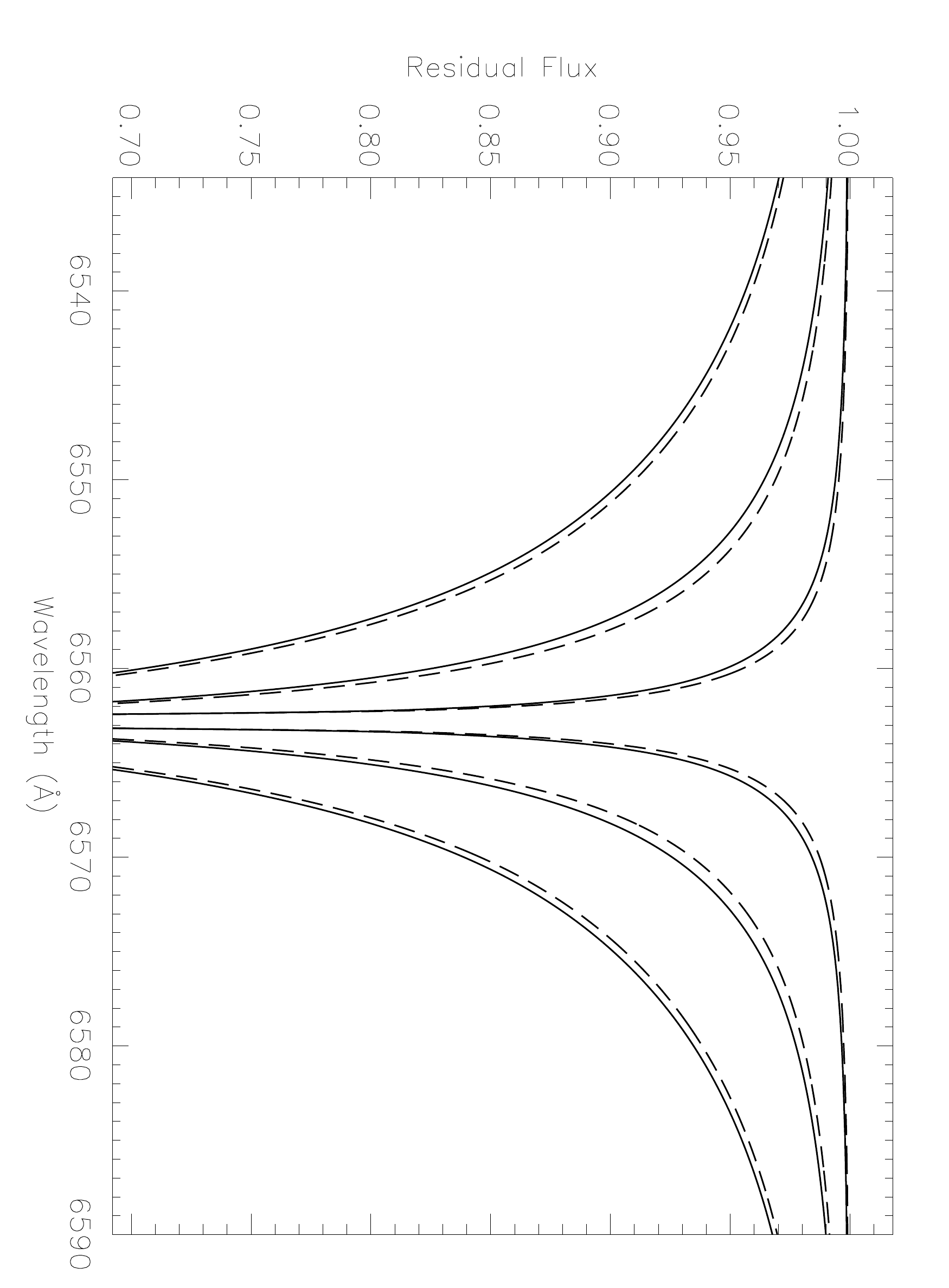}}
    \caption{Sensitivity of computed profiles for H$\alpha$ with the stellar
effective temperature ($T_{\mathrm{eff}}$~=~5000, 6000, 7000~K), for an atmosphere
with solar surface gravity and composition. Stronger
lines correspond to warmer atmospheres. The differences between solid and dashed
profiles illustrate variations between two theoretical calculations of the
contribution of H-H collisions. 
Image reproduced with permission from \cite{2000A&A...363.1091B}, copyright by ESO.}
    \label{balmer}
\end{figure}}

The surface gravity of a star is the parameter that has the
smallest effect on the spectrum. This is unfortunate 
to constrain it from the spectrum itself, but at the same time 
it means that this parameter will have a limited impact 
on the derived abundances. As mentioned earlier, there are
cases in which we can use external data to constrain 
surface gravity, but until the Gaia catalog is 
available this will not apply to many stars. 

Using the spectrum,
gravity can be constrained taking advantage of the way the 
electron pressure enters in Saha's equation~(\ref{saha}): 
at a given temperature,
the lower the electron density, the higher the ionization. Such
effect is readily noticeable in cool stars on the H bound-free ionization
edges, e.g., in the Balmer jump in the optical. The wings of strong
metal lines are also very useful to constrain gravity. As surface
gravity increases, so does the effect of broadening by collisions
between the absorber and H (and partly He) atoms, enhancing the wings 
of metal lines, as illustrated in Figure~\ref{MgIb} for one of the
lines of the Mg\,{\sc i}\textit{b} triplet in the spectrum of the
metal-poor star BD+17~4708.

\epubtkImage{}{
  \begin{figure}[htb]
    \centerline{\includegraphics[width=1.0\textwidth]{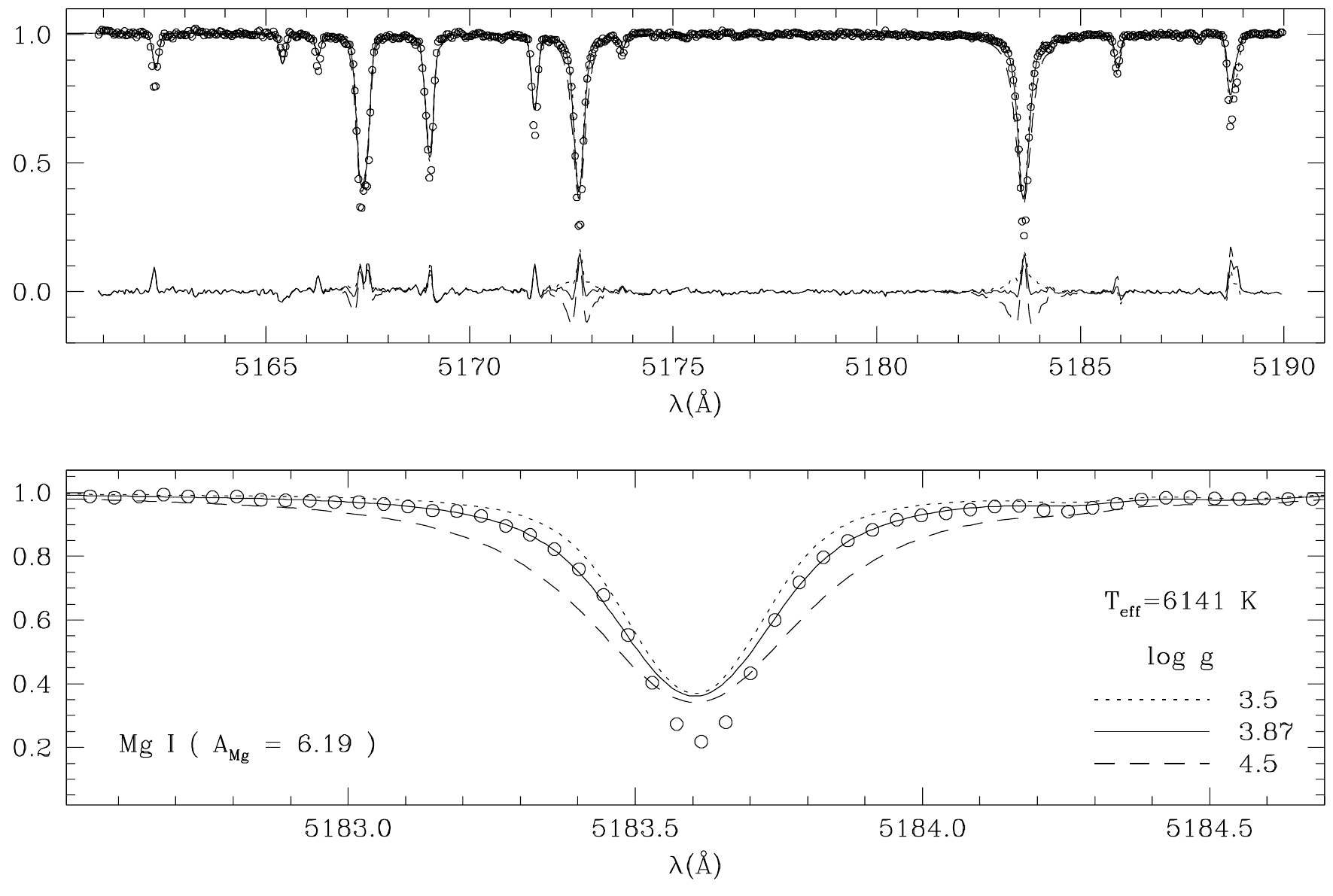}}
    \caption{The top panel shows part of the spectrum of the metal-poor star BD+17~4708, while
the bottom panel provides a close view of one of the Mg\,{\sc i}\textit{b} lines in the spectrum
of the star compared with calculations for models with a range in surface gravity. 
The analysis focuses on the wings, since the line core is not matched due to
departures from LTE.
Image reproduced with permission from \cite{2006A&A...459..613R}, copyright by ESO.}
    \label{MgIb}
\end{figure}}

The overall metallicity, usually approximated by the
iron abundance, is derived, like any other abundance,
from lines of weak-to-moderate strength. This is either done
from equivalent widths or line profiles, but this choice will be further 
discussed in the next section.

\subsection{Determining abundances from spectra}
\label{abundances}

Once we have decided on the best set of atmospheric parameters
for a star which, we recall, are likely to be revised during the analysis, we can proceed 
calculating, or obtaining from a data base, an appropriate model
atmosphere. Of course, if some or all of the atmospheric parameters 
are constrained from observed line strengths, as opposed to
other external information such as photometry, spectrophotometry,
or astrometry, we will need the model atmospheres earlier in the process.
Any theoretical fluxes to be compared to observations, 
i.e., photometry or spectrophotometry, also require model atmospheres,
but unlike high-resolution spectra, these fluxes are usually 
available together with the libraries of model atmospheres, 
and one does not need to calculate them in most instances.

Depending on the type of star, the accuracy
being sought, more mundane factors such as access to 
computing resources and human collaborators, we should decide on the type 
of model atmosphere to adopt, and how to secure it. We have already mentioned
the most widely used possibilities in Section~\ref{models}.

The Kurucz codes have been used in modern operating systems,
and are publicly available (e.g., from Castelli's 
site%
\epubtkFootnote{\url{http://wwwuser.oat.ts.astro.it/atmos/} and 
\url{http://wwwuser.oat.ts.astro.it/castelli/}}). 
The \textsc{MARCS} code is not, although 
custom-made models are kindly provided by the core \textsc{MARCS}
community -- mostly residents or former residents 
of the Swedish town of Uppsala. There are many different interpolation 
tools that have been used in the literature, although few are 
publicly available and documented. An interpolation
code for \textsc{MARCS} models is available from the \textsc{MARCS} site, and
others for Kurucz's models are available from A.~McWilliam 
and from me.%
\epubtkFootnote{See \textsc{kmod.pro} at \url{http://www.as.utexas.edu/~hebe/stools/}}

As mentioned earlier, in addition to a model atmosphere, 
deriving abundances requires
basic data associated with the atomic or molecular transitions
we plan on using as diagnostics: wavelengths, transition probabilities, the 
energy of the lower level connected by the transition, and 
damping constants. With these inputs, we have all we need to 
calculate a model spectrum, since other auxiliary data are 
typically included in radiative transfer codes. Once we are able
to compute model spectra, we can iteratively compare those with
observations to infer the chemical compositions of stars. 
Section~\ref{tools} is devoted to tools commonly used for 
these tasks.

\subsection{Tools for abundance determination}
\label{tools}

As outlined above, the process of abundance determination requires
a certain degree of iteration. At the very least, when the atmospheric
parameters are well constrained, and the overall metal content of the star 
is known in advance, we need to iterate the last step of calculating
a synthetic spectra to arrive at the chemical composition that best matches 
 the observations. 

Traditionally, for the analysis of a small number of spectra, 
iterations are performed under the careful control of a person.
The basic workhorse software is a radiative transfer code,
that predicts what the observed spectrum would look like for
a given model atmosphere and chemical composition. We mention the
most popular choices (there are many more!) in Section~\ref{interactive}.
When the number of spectra to analyze is very large, 
a manual analysis using interactive software is
no longer possible, and more powerful tools become necessary. These
are the subject of Section~\ref{automated}.

\subsubsection{Interactive tools}
\label{interactive}

The fundamental tool for chemical analysis 
is a spectral synthesis code. For a fixed set of input data:
model atmosphere, and atomic and molecular line and continuum opacities,
the code can be used to predict spectra for different abundances
to identify the value that matches observations -- one or an array of
absorption lines. Some of the most commonly used spectral synthesis codes publicly 
available have already been mentioned in Section~\ref{radiative_transfer}: these are \textsc{MOOG}%
\epubtkFootnote{\url{http://www.as.utexas.edu/~chris/moog.html}}
\citep{1974PhDT.......316S}, 
\textsc{Synspec}%
\epubtkFootnote{\url{http://nova.astro.umd.edu/Synspec49/synspec.html}}
by Hubeny \& Lanz, 
\textsc{Turbospectrum}%
\epubtkFootnote{\url{http://www.pages-perso-bertrand-plez.univ-montp2.fr/}}
by \cite{2012ascl.soft05004P} \citep[see also][]{1998A&A...330.1109A}.
Many more codes are available but less frequently used, for example
\textsc{MISS}
by \cite{1998ApJ...502..951A} \citep[see also][]{1992ApJ...398..375R},
\textsc{RH}
by Uitenbroek\epubtkFootnote{\url{http://www4.nso.edu/staff/uitenbr/rh.html}},
\textsc{ASSET} by
\cite{2009AIPC.1171...73K}, 
or \textsc{SPECTRUM}%
\epubtkFootnote{\url{http://www.appstate.edu/~grayro/spectrum/spectrum.html}}
by R.\ O.\ Gray. 

Using line equivalent widths (the total absorption associated with
a spectral line) is operationally convenient (one number per transition 
instead of an array of fluxes) and makes the results independent 
from macro-turbulence
and rotational broadening. For those reasons, despite it represents a
loss of information, equivalent widths are still used in many studies.
This requires measuring line equivalent widths in both observed
and model spectra. One of the software packages commonly used to measure manually
equivalent widths is the \emph{splot} package within \textsc{IRAF}.%
\epubtkFootnote{IRAF (Image Reduction 
and Analysis Facility) is distributed by the National Optical Astronomy Observatory, which 
is operated by the Association of Universities for Research in Astronomy (AURA) under 
cooperative agreement with the US National Science Foundation.} 
Codes such as \textsc{MOOG} can be used to provide abundances directly from 
equivalent widths (see Section~ref{automated}). 
This modus operandi is highly effective for
processing large numbers of transitions for many stars, and it has
been used with success in numerous projects 
\citep[see, e.g.,][to mention a few examples]{2003MNRAS.340..304R,
2003A&A...410..527B,
2004A&A...415.1153S,
2010ApJ...724L.104F,
2015A&A...579A..52N}.

\subsubsection{Automated tools}
\label{automated}

There are several codes that automate the measurement of 
equivalent widths. The most used ones are \textsc{ARES} \citep{2007A&A...469..783S} 
and \textsc{DAOSPEC} \citep{2008PASP..120.1332S,2014A&A...562A..10C}.
As already mentioned, the derivation of abundances from observed equivalent widths 
has been automated for a long time in codes such as \textsc{MOOG} or those
in the Uppsala package. The automation of the task of directly fitting spectra
with models has a more recent history.

\textsc{Spectroscopy Made Easy}%
\epubtkFootnote{\url{http://www.stsci.edu/~valenti/sme.html}}
by \cite{1996A&AS..118..595V}
is a package for fitting spectra by wrapping
an optimization algorithm around a spectral synthesis code. 
\textsc{ULySS}%
\epubtkFootnote{\url{http://ulyss.univ-lyon1.fr/}}
by \cite{2009A&A...501.1269K}
is also spreading in use. \textsc{FERRE}%
\epubtkFootnote{\url{http://hebe.as.utexas.edu/ferre}}
by \cite{2006ApJ...636..804A}
includes several algorithms and is optimized for dealing with large samples.
Another option is 
\textsc{VWA}%
\epubtkFootnote{\url{https://sites.google.com/site/vikingpowersoftware/}} 
by \cite{2010A&A...519A..51B}.
Many more packages of general application have been employed in the literature, but 
they are not publicly available, e.g., 
\textsc{MATISSE} 
by \cite{2006MNRAS.370..141R}, 
\textsc{TGMET}%
\epubtkFootnote{\url{http://www.obs-hp.fr/guide/elodie/tgmet-doc/TGMET1.HTML}}
by \cite{1998A&A...338..151K}, 
or \textsc{MyGIsFOS}
by \cite{2013arXiv1311.5566S}. 
Some packages have been developed
for particular instruments or projects, such as various parts of the SEGUE Stellar
Parameter Pipeline 
\citep{2008AJ....136.2022L,2008AJ....136.2050L,2008AJ....136.2070A, 
2011AJ....141...90L}, 
or the original RAVE analysis pipeline
\citep[e.g.,][]{2011AJ....142..193B,2011AJ....141..187S}.

Although these tools are very useful to deal with large samples, 
one should avoid using them blindly. There is no substitute for
understanding which spectral features can be modeled accurately
and which ones cannot. The importance of checking thoroughly, using well-studied
stars, the performance of an automated tool on a particular data set, 
cannot be overemphasized.

\newpage


\section{Observations}
\label{data}

There is a wide variety of collections of spectra readily available
on the internet. This includes libraries with typically high-quality data for
specific sets of stars, and data repositories for large surveys.
We describe the most significant ones below, as well as the most exciting 
instruments now in operation or being planned for the future.

\subsection{Libraries}
\label{libraries}

If you are interested in a bright star, its high-resolution
optical spectrum may be available in the 
Elodie archive%
\epubtkFootnote{\url{http://atlas.obs-hp.fr/elodie/}}
\citep{2004PASP..116..693M}, 
S\super{4}N%
\epubtkFootnote{\url{http://hebe.as.utexas.edu/s4n/}}
\citep{2004A&A...420..183A}, 
the UVES-Paranal library%
\epubtkFootnote{\url{https://www.eso.org/sci/observing/tools/uvespop/interface.html}}
\citep{2005IAUS..228..261J}, 
or the Nearby Stars Project%
\epubtkFootnote{\url{http://bifrost.cwru.edu/NStars/}}
\citep{2005AJ....129.1063L}. 
If the spectrum has been ever obtained with HST, you will find it
in MAST,%
\epubtkFootnote{\url{https://archive.stsci.edu/}}
and if it has been 
secured in recent times with ESO instruments, 
there is a good chance it is publicly 
available in the ESO archive%
\epubtkFootnote{\url{http://archive.eso.org/cms.html}}, 
although you may 
need to reduce the data yourself. Other observatories have their 
own archives, and very many are now becoming publicly available, 
but they are too many to list here, and their
reservoir of public high-dispersion spectra is anyway limited.
A new library under development at Leiden is based on
observations with X-Shooter, offering a wide spectral
coverage at medium-high resolving power \citep{2012ASInC...6...13C}. 

Regarding spectrophotometry, it is 
worthwhile to highlight the Indo-US library%
\epubtkFootnote{\url{http://www.noao.edu/cflib/}}
by \cite{2004ApJS..152..251V}, 
and MILES%
\epubtkFootnote{\url{http://www.iac.es/proyecto/miles/pages/stellar-libraries/miles-library.php}} 
by \cite[][and references therein]{2011A&A...532A..95F}
from the ground, and the STIS Next Generation 
Spectral Library%
\epubtkFootnote{\url{https://archive.stsci.edu/prepds/stisngsl/}} by
\cite{2006hstc.conf..209G} and \cite{2007IAUS..241...95H} 
from space. The ESA mission Gaia should bring accurate 
parallaxes and spectrophotometry (with very low resolution though) 
for $10^9$ stars before the end of the decade 
\citep{2008IAUS..248..217L}, 
but we will come back to this important observatory in Section~\ref{future_surveys}.

\subsection{Ongoing projects}
\label{ongoing_surveys}

We live in a time when instruments with high multiplexing 
capabilities are revolutionizing the study of the chemical compositions of
stars in the Milky Way and other nearby galaxies. For the most part,
the distinction between instruments and surveys is blurred,
since these instruments serve mainly
a single or a small number of projects.

For the sake of limiting the size of this article, we will concentrate
on the largest projects and the instruments with the largest multiplexing
capabilities currently in operation, and which are being massively used
for the acquisition of stellar spectra that have become or will soon
become public, at the expense of leaving out excellent instruments 
that do not meet those conditions.

The Sloan Digital Sky Survey \citep[SDSS;][]{2000AJ....120.1579Y}
is the most dramatic example of a modern survey making efficient use of a
suite of instruments. In addition
to five-band photoelectric photometry of a third of the sky, 
the SDSS has obtained 
$R=2000$ spectra for several million galaxies and about a 
million stars since it started operations in 2000. The project
uses a dedicated 2.5-m telescope 
\citep{2006AJ....131.2332G}
with a field-of-view of 7 square
degrees. Originally equipped with two double-arm spectrographs 
\citep{2013AJ....146...32S}, 
fed by 640 fibers, and subsequently upgraded to 1000, providing
a wide spectral coverage (360\,--\,1000~nm approximately) the project 
concentrates on stars in the magnitude range $14 < V < 21$.
These data have been used for studying the structure of the Milky
Way disk and halo in a series of papers
\citep[see, e.g.,][]{2006ApJ...636..804A,2009ApJ...703.2177S, 
2010ApJ...716....1B,2011ApJ...734...49S, 
2011ApJ...738..187L,2012ApJ...752...51C, 
2012ApJ...749...77S,2012ApJ...761..160S,
2015A&A...577A..81F,
2014A&A...567A.106L}, 
but many more studies are
to be expected as observations continue (at least to 2020).
SDSS releases their data publicly every one or two years, a move
that has been decisive in making it one of the most scientifically
productive observatories in the world.

A few years ago, SDSS added a new instrument optimized for the observation
of stars in the H-band ($1.5\mbox{\,--\,}1.7\mathrm{\ \mu\,m}$) at high $R=22\,500$. The Apache Point
Galactic Evolution Experiment (APOGEE) has collected so far nearly half a million
spectra for some 100\,000 stars in the Milky Way 
\citep[see][]{2015arXiv150104110H,2015arXiv150905420M}
and results are already appearing on various subjects
\citep[see, e.g.,][]{2013AJ....146..156D,2013ApJ...777L..13M,
2013ApJ...777L...1F,2013AJ....146..133M, 
2013AJ....146...81Z,2013ApJ...767L...9G,2012ApJ...759..131B,
2014ApJ...796...38N,2015ApJ...808..132H}.
The project will expand this collection of spectra by a factor of several between 2014-2020, 
including observations from Las Campanas, accessing the Magellanic clouds, 
parts of the disk not available from the North, and getting
a better perspective of the central parts of the Milky Way.
In the SDSS tradition, APOGEE's spectra and derived data products are
made public on a regular basis.

The Radial Velocity Experiment (RAVE) 
\citep[see, e.g.,][and references therein]{2013AJ....146..134K}, 
already done with the data taking,
has targeted stars with brighter magnitudes ($5 < V < 14$) and therefore 
located at closer distances, with higher resolution $R=8500$ and a narrower spectral
range (837\,--\,874~nm) than SDSS. The results from the survey are already coming out,
\citep[among their most recent publications see, 
e.g.][]{2013AJ....146..134K,2014A&A...562A..91P, 
2014A&A...562A..54C,2013MNRAS.436.3231K,2013ApJ...778...86K, 
2013MNRAS.436..101W}
and the associated spectra for about half a million stars 
will probably become public sometime in the future.

The Large Area Multi Object fiber Spectroscopic Telescope (LAMOST) 
resembles in many respects SDSS but with 4000 fibers, 
an original telescope with an effective aperture between 3.6\,--\,5.9~m, 
and a robotic fiber positioner. After
some difficulties the project is now starting to show results 
\citep[see, e.g.,][]{2014AJ....147...33Y,2014NewA...26...72Y, 
2013AJ....146...82R,2013AJ....146...34Z,2013AJ....145..169Z,2015RAA....15.1089L},
and a lot more is envisioned to come out in coming years.

The Gaia-ESO Survey (GES) is a massive ESO public survey that is using
300 nights on the Very-Large Telescope to obtain 
high-resolution spectra for about 100,000 stars
\citep{2012Msngr.147...25G}, 
thought out to complement the observations from Gaia (see Section~\ref{future_surveys}).
About 90\% of the observations are made at a resolving power of about 15,000
with the GIRAFFE spectrograph \citep{1999JAF....60...19H,2000SPIE.4008..467B}, 
and the remainder at higher resolution with the
cross-dispersed UVES spectrograph \citep{1992ESOC...42..581D,1999JAF....60...19H}. 
It involves a large European collaboration, and has two distinct
parts, one focusing on field stars and a second one devoted to clusters. 
The observations started at the end of 2012, the raw spectra
are quickly made public on the ESO archive, and the advanced data products
(fully reduced spectra, radial velocities and stellar parameters
including abundances) are following on a fast-paced schedule.
There are already plenty of examples of science results from this project
\cite[e.g.,][]{2015A&A...580A..85M,2015A&A...575L..12L,2015A&A...574L...7S}.

The GALactic Archaeology with HERMES project (GALAH) 
\citep[see][]{2013AAS...22123406Z} 
expands on the idea of chemical tagging and producing a chemical map of stars
in the Galaxy to provide optical high-resolution spectroscopy ($R=28\,000$)
for over a million stars using the 4-m AAT telescope in Australia. 
The HERMES instrument was commissioned at the end of 2013, and the first 
results are eagerly expected.

The Hobby-Eberly Telescope Dark Energy Experiment 
\citep[HETDEX;][]{2012AAS...21942401H}, 
although conceived mainly as a cosmology project, leaves no star behind, 
at least of those that fall on their Integral-Field Units during
the course of the experiment, which last five years. These spectra
will be deep ($V<22$) and reach into the UV (350\,--\,550~nm), although 
of low resolution ($R=700$), with the 
particular plus of having no selection biases (like Gaia). As of this
writing the observations are a few months from starting.

\subsection{In the future}
\label{future_surveys}

If current and ongoing surveys are tremendously exciting, there are 
instruments in the design and construction phase that will make
the future even more thrilling. Having in mind that this is 
a live review, and accepting the task of coming back to complete
information as time progresses, the following paragraphs
will only mention projects that are subjectively perceived by this
writer as the most promising ones.

Gaia has long been declared as the ultimate survey of the Milky Way.
There is no doubt that Gaia's capabilities are impressive. 
The satellite was launched in December 2013, and the first catalog
is expected in 2016. This ESA cornerstone mission is building
a full-sky catalog of the stars in the Milky Way with a unique suite
of instruments that provide exquisite astrometry, and 
 (spectro-)photometry for about $10^9$ stars, and high-resolution
spectroscopy (847\,--\,874~nm) for about $10^8$ stars. Rather than providing
all the details on Gaia, I will simply refer the reader to 
\cite{2008IAUS..248..217L}, but not without noting that all other spectroscopic
efforts are largely complementary to Gaia, since its high-resolution 
spectrograph is only effective to provide radial velocities for
a small fraction of the sample, due to the short integration times and the implied 
low signal-to-noise ratio of the spectra.

The Dark Energy Spectral Instrument \citep[DESI;][]{2013arXiv1308.0847L} 
is an ambitious project conceived as a natural follow-up of 
the Baryon Oscillations Spectroscopic Survey 
\citep[BOSS;][]{2007AAS...21113229S,2011AJ....142...72E}, 
part of the SDSS. Improving on telescope area (from the 2.5-m SDSS telescope to the
4-m Mayall telescope at Kitt Peak), doubling spectral resolution,
 enhancing multiplexing capabilities (from 1000 to 5000 fibers, 
from plug-plates to a robotic positioner), 
all without losing field of view and synoptic operations. 
Like BOSS and the follow-up eBOSS project, DESI will mainly measure redshifts 
to identify the imprint left by primordial 
baryonic acoustic oscillations on large scale structure,
but there is no doubt that this instrument will also make an important
contribution for Galactic studies of stars.

In addition to DESI, there are other instruments planned
for 4-m-class telescopes. These instruments have somewhat more modest
multiplexing capabilities but differ from DESI 
in that they provide various spectral settings, allowing higher resolution 
observations, and therefore will enable stellar 
science that DESI cannot do, and smaller, more targeted projects. These
are WEAVE for the 4.2-m WHT in La Palma \citep{2014SPIE.9147E..0LD}, or
4MOST for the 4-m VISTA
at Paranal \citep{2014SPIE.9147E..0MD}. Yet another high-multiplexing 
spectrograph with the ability of reaching a resolving power of $R=20\,000$
in the infrared for up to $\sim$~1000 targets simultaneously, 
MOONS, is projected for the 8-m ESO VLT 
\citep{2014SPIE.9147E..0NC,2014SPIE.9147E..2CO}.

One can imagine the ultimate spectroscopic experiment, let's call it
the Final Uniform Spectroscopic Survey, using a Field Ultra-dense 
Survey Spectrograph (FUSS), a natural endpoint of
all the surveys described above, providing radial velocities and chemistry
to complement Gaia's astrometry and photometry. This would consist of a 
pair of large-field large-aperture telescopes (one per hemisphere), perhaps 6 to 8-m-class, 
resembling the LSST, with massive integral field units (IFU) covering 
their entire field of view \'a la VIRUS (the HETDEX spectrograph), 
surrounded by a farm with tens of thousands of compact fiber-fed spectrographs.
Over the course of the project, probably about a decade, 
these telescopes would scan the entire sky
once, getting spectra of everything down to $V$~=~20~mag. This project has
not yet been seriously proposed, but I would be surprised if it is not
started by 2025!

\subsection{Examples of recent applications}

The analysis of stellar spectra to derive
chemical compositions of stars has led, in the last decade,
to vast progress across fields in astrophysics. With the aim
of highlighting practical applications of stellar photospheric 
abundances, a series of topics that are the subject 
of active research are briefly presented below. These are only
a few examples, and other writers would surely choose other topics.

%

%

\subsubsection{The Galactic disk}

Stars in the Galactic disk were found to split nicely into two
components with densities falling exponentially from the plane 
and scale heights of 
about 0.3 (thin disk) and 1~kpc \citep[thick disk][]{1983MNRAS.202.1025G}.
The spectroscopic studies of these populations carried out in the
last decade have also shown a split in their $\alpha$-element
(O, Mg, Si, Ca, Ti) to iron abundance ratios, as well as in
age, with stars in the thick disk being fairly old ($>$~8~Gyr)
and those in the thin disk younger, between 1\,--\,8~Gyr 
\citep[see, e.g.,][]{1998A&A...338..161F,
2003A&A...410..527B, 
2006MNRAS.367.1329R,
2006ApJ...636..804A,
2013A&A...554A..44A,
2014ApJ...796...38N,
2014A&A...564A.115A,
2015MNRAS.450.2874R}.

Despite the rich information available from spectroscopic data, 
it is yet unclear how this two-component structure
came together, and even whether the two disks are intimately connected or not. 
It is recognized that gas accretion from early galactic neighbors 
and radial migration of stars in the disk must have
played a role, but the sequence of events that gave rise to
the present-day Galactic disk has so far evaded our understanding. 
This is an area where the data flow from Gaia and the spectroscopic
surveys mentioned above, 
hand by hand with models and simulations,
will likely be of great help in the following years.

\subsubsection{Globular clusters}

For the most part of the 20th century, globular clusters
were considered groups of stars sharing age and chemical
composition. Unlike Galactic (open) clusters, globulars
are old and metal-poor, associated with the Milky Way halo (or thick disk), 
and as such are excellent probes to study the early evolution of the Galaxy.
Then we came to realize that was not true.
Star to star variations in some elements became apparent.
Sodium to oxygen or magnesium to aluminum correlations. 
Even worse, space-based photometric studies revealed
multiple sequences, suggestive of different helium fractions, 
in the most massive clusters \citep[see, e.g.,][]{2004ApJ...612L..25N}.

The details of how globular clusters formed remain elusive, 
but it seems clear that some of the stars in the clusters
formed from gas that has been polluted by previous cluster
stars \citep{2004ARA&A..42..385G}. 
Furthermore, at least some of the most massive
globular clusters in the Milky Way appear to be the remnants
of small galaxies that ended up torn apart in the Galactic potential 
\citep{2003A&A...405..577B}.

\subsubsection{The most metal-poor stars}

Since the chemical composition of the Universe 
after the primordial nucleosynthesis following the 
Big Bang was basically hydrogen and helium, the 
overall lack of heavier elements in a star can be interpreted as 
a sign of a primitive composition and a very old age. 
Observational cosmology has traditionally targeted galaxies 
at high redshift, but the objects with the most primitive
compositions have been found right here in the Milky Way. These
are extremely metal-poor stars with iron-to-hydrogen ratios
orders of magnitude lower than the Sun: up to 7 orders of magnitude in the case
of the star recently discovered by \cite{2014Natur.506..463K}.

The relative numbers of stars with different metallicities
in the halo constrain the star formation history and initial 
mass function in the early Milky Way. Furthermore, the oldest
stars in the Galaxy can inform us about what happened 
in the early Universe.

For stars to form, the gas in the interstellar medium needs
to cool down for gravity to win over pressure.
At very low metallicity, the lack of lines from metals
makes it very difficult for gas to radiate and cool down.
As a result, massive stars form, but no low-mass stars, 
producing a mass distribution shifted to higher
masses than what we observe today in the disk of the Milky
Way \citep{2004ARA&A..42...79B}.
The discovery a few years ago by \cite{2012A&A...542A..51C}
of a dwarf star in the halo with an overall metallicity of [Fe/H]$\simeq -4$
sets an upper limit to where that threshold can be, 
constraining theories of star formation.

\subsubsection{Solar analogs}

Uncertainties in atomic and molecular data, especially oscillator strengths, 
directly propagate into the derived abundances from stellar spectra. That
applies to absolute abundances. However, in 
a differential analysis between two stars that are very similar, 
systematic errors cancel out.

Differential studies between the Sun and similar stars can render a 
precision in the derivation of atmospheric parameters at the level of a few K
in $\mathrm{T}_{\mathrm{eff}}$ and 0.01~dex in $\log g$, which are also accurate to that 
level, since the absolute surface temperature and gravity of the Sun are known
even better. In this fashion, relative 
abundances can be derived to better than 0.01~dex in [Fe/H]
\cite{2015A&A...579A..52N}, or more than 10 times better 
than absolute abundances.

This level of precision opens new opportunities for discovery in astrophysics.
Detailed studies of nearby stars with parameters very close to solar 
have shown that there are subtle differences in composition between 
most of them and the Sun \citep{2009ApJ...704L..66M,2009A&A...508L..17R},
which may be related to the sequestration in rocks or rocky planets of 
refractory elements that do not make it into the star. Needless to say,
we look forward to future applications of this technique to other
types of stars.

\section{Reflections and Summary}

The analysis of the chemical compositions of stars evolved
from early experiments by Bunsen, Kirchhoff and Fraunhofer in the 19th
century into a quantitative field through the 
theoretical foundations provided by Schwarzschild, Eddington, Milne, and others.
The LTE numerical codes for computing model atmospheres developed
in the 1970s and 80s, mainly by Gustafsson and Kurucz, 
set the standard for modern analyses of AFGKM-type stars, 
while NLTE codes were developed for warmer ones.
Atomic and molecular data, at a modest pace, has been perhaps the
most significant improvement in modeling late-type stellar spectra over the last 
two decades, with the exception of the development of hydrodynamical 
models, which represent a dramatic advancement 
in our ability to match in close detail spectral line shapes, even 
though their overall impact on the inferred abundances is limited.

The rate at which spectroscopic observations are gathered has been 
revolutionized, and a few individual projects have obtained more spectra in the
last few years than the accumulated data from previous history.
These projects are providing vast and homogeneous data sets,
which have not been fully explored, while new and larger projects
are appearing in the scene. In response, spectral analysis software
is being transformed from interactive tools to industrial chain processing. 

The extraordinary new observations soon to become available from the Gaia mission,
and the constraints on atmospheric parameters for individual stars that they
will provide, 
will help enormously to uncover the deficiencies of standard
analyses, driving improvements in modeling spectra. The new spectroscopic 
data from large ongoing and upcoming ground-based surveys will slowly but
surely enlighten our understanding of the formation and evolution of 
the Milky Way, and as result, of galaxy formation and evolution in general.

These large spectroscopic efforts are producing large data bases of 
atmospheric parameters and chemical abundances, which in order to materialize into
knowledge will need to be confronted with models of galaxies and
their chemical evolution. At the same time, the making of such models is 
informed by the results of observational efforts. Hopefully, the process
will, over the next decade, converge to provide us with some clear quantitative
statements on many of the pending questions. Among the highest priority 
issues, we need to find out how many of the 
Milky Way stars were formed in situ and how many come from smaller galaxies
that merged, what are the time scales and formation processes 
for each of the main stellar populations in the galaxy, and how they relate.

Massive observational projects are providing us with
a detailed census of a significant fraction of the 
Milky Way stars and their fundamental parameters. This information will allow
progress on many other fields aside from galaxy evolution. To cite a few examples,
the information will shed light on stellar evolution, extrasolar planetary
systems and their relation to the chemistry of their hosting stars, the
structure of the interstellar medium, or star formation and stellar dynamics.


\section*{Acknowledgements}
\label{sec:acknowledgements}

Thanks are due to those that set the foundations for the work we are 
currently doing, and of course to my colleagues for their partnership 
and patience. I am in debt to Paul Barklem, Nicolas Grevesse, Ivan Hubeny, 
Basilio Ruiz Cobo, and an anonymous referee, for valuable comments and suggestion
on a draft of this article. I would like to request readers that find
mistakes or omissions to please send those to me and they will be
fixed!

\newpage



\bibliography{refs}

\end{document}